\documentclass[a4paper,12pt]{article}
\usepackage[top=3cm,left=3cm,bottom=2cm,right=2cm]{geometry}
\usepackage[utf8]{inputenc}
\usepackage[T1]{fontenc}
\usepackage[brazil]{babel}
\usepackage{newtxtext,newtxmath}
\usepackage{graphicx}
\usepackage{csquotes}
\usepackage[giveninits]{biblatex}
\usepackage[euler]{textgreek}
\usepackage{indentfirst}
\usepackage[labelfont=bf,labelsep=endash,font=footnotesize]{caption}
\usepackage{subcaption}
\usepackage{fancyhdr}
\newenvironment{citacao}{
  \begin{list}{}{
    \setlength{\leftmargin}{4cm} 
    \setlength{\topsep}{18pt}    
    \setlength{\parsep}{0pt}
    \setlength{\itemsep}{0pt}
    \linespread{1.0}             
    \fontsize{10}{12}\selectfont 
  }
  \item[]
}{
  \end{list}
}
\newcommand{\source}[1]{%
  \caption*{\fontsize{10}{10}\selectfont Fonte: #1}%
}

\title{Astrofísica Observacional e Ensino de Física Moderna: Uma Abordagem Conceitual da Espectroscopia Astronômica}
\author{
  Brunno Pleffken Hosti\footnote{Graduado em Física pela Pontifícia Universidade Católica do Paraná – brunno.hosti@pucpr.edu.br}
  \and
  Ângela Cristina Cararo\footnote{Professora adjunta da Pontifícia Universidade Católica do Paraná – angela.cararo@pucpr.br}
}
\date{}

\begin{document}

\maketitle

\bigskip

\begin{center}
  \textbf{RESUMO}
\end{center}

\noindent{A formação continuada de professores em ciências exatas é fundamental para a qualidade do ensino, especialmente ao se abordar a física moderna. Apesar disso, a didática desses temas frequentemente se apoia em modelos teóricos que podem parecer abstratos e distantes das aplicações práticas. Nesse contexto, a pesquisa em astrofísica fornece muitas percepções valiosas quanto à natureza da luz e suas propriedades fundamentais, tais como o espectro contínuo e discreto, a radiação de corpo negro e os orbitais atômicos segundo o modelo atômico de Bohr. Este artigo, destinado a professores licenciados em Física do ensino médio e ensino superior, visa examinar as particularidades dos espectros de emissão e absorção de diversos tipos de objetos astronômicos e demonstrar como a espectroscopia é aplicada de maneira prática na pesquisa científica em astrofísica. Nesta abordagem, o presente estudo apresenta de forma conceitual como os astrofísicos, por meio da medição dos espectros da luz, determinam a composição, as propriedades físicas, a origem e a evolução dos diversos corpos celestes e, por extensão, do universo como um todo. Ao compreender não só a teoria, mas também as aplicações diretas da espectroscopia astronômica, os professores estarão mais preparados para guiar seus alunos, mostrando o verdadeiro valor da física moderna no mundo real.}

\bigskip

\noindent{\textbf{Palavras-chave:} astrofísica observacional; espectroscopia; ensino de física moderna.}

\section{Introdução}

A física moderna fornece os fundamentos para compreender os princípios que regem o comportamento do universo. Ao aprender sobre a nova Física desenvolvida a partir do século XX, os estudantes podem desenvolver uma compreensão mais profunda da natureza da matéria, da energia e das forças fundamentais que moldam nosso mundo \cite{pagliarini:almeida}. Como enfatizam \textcite{silva:almeida}, os temas dessa área são conhecidos por envolver conceitos muito abstratos e contra-intuitivos, mas, ao lidar com essas ideias, os estudantes desenvolvem a capacidade de raciocínio lógico e de analisar problemas complexos, habilidades que não se limitam ao campo da Física, mas para diversas outras áreas da vida.

Para estudantes que desejam seguir carreira nas áreas das ciências exatas, engenharia ou áreas afins, uma base sólida em física moderna é crucial, pois áreas de estudo como a mecânica quântica e a física atômica e molecular fornecem a base para estudos e pesquisas científicas de alto nível e com grande impacto na sociedade. O acelerador de partículas Sirius, localizado em Campinas/SP, é um exemplo do efeito direto da física moderna no desenvolvimento da ciência nacional. A introdução desses conceitos em fase escolar pode contribuir para uma melhor preparação dos estudantes para os desafios que enfrentarão no ensino superior e na pós-graduação, mas, para atingir esse objetivo, é necessário um preparo adequado dos professores, a fim de que estejam capacitados a ministrar uma aula de qualidade \cite{silva:almeida}.

Em uma sociedade cada vez mais impulsionada pela ciência e tecnologia, é essencial que os cidadãos tenham uma compreensão básica dos conceitos fundamentais da Física. Muitas questões de relevância social, como a busca por novos materiais, a produção de energia limpa e avanços na medicina, são influenciadas por esses princípios — a superação de tabus acerca da energia nuclear, como aborda \textcite{tosi}, para a substituição da matriz energética fóssil é um exemplo do impacto direto dos conhecimentos de Física na sociedade ao nível global. Ao educar os estudantes sobre eles, nós os capacitamos a tomar decisões informadas e a participar de discussões sobre esses tópicos importantes. Conforme \textcite{sanches}:

\begin{citacao}
  A importância da introdução da Física do século XX no currículo é também reconhecida, no sentido de construir no estudante a consciência de sua responsabilidade social e ética. Desse modo, é necessário promover competências para que o jovem seja capaz de avaliar a veracidade de informações ou para emitir opiniões de juízos de valor em relação a situações em que os aspectos físicos são relevantes \cite{sanches}.
\end{citacao}

A importância do conhecimento dos fenômenos físicos para o desenvolvimento da tecnologia moderna é, também, enfatizada por \textcite{carvalho}:

\begin{citacao}
  Podemos destacar dentre os motivos aqueles que permitem que os alunos dialoguem com os fenômenos físicos que estão por trás do funcionamento de aparelhos que, atualmente, são utilizados de forma corriqueira no dia-a-dia da maioria das pessoas, fato, aliás, que torna o assunto bastante interessante \cite[p.~2]{carvalho}.
\end{citacao}

A física moderna e contemporânea, juntamente com a astronomia e a astrofísica, é um campo que expande os limites de nossa compreensão do universo. Apresentá-la a estudantes do ensino médio pode despertar curiosidade e admiração, inspirando alguns a seguir carreiras científicas e contribuir para o avanço do conhecimento e da tecnologia. Mesmo para estudantes que não buscam seguir carreira científica, o contato com a física moderna pode cultivar um sentimento de admiração e apreço pelo mundo natural. Muitos trabalhos desenvolvidos destacam a importância do ensino de Ciências da Natureza em fase escolar como incentivo para os jovens seguirem a carreira na pesquisa. Como justificativa, destaca-se o trabalho de \textcite{ostermann} acerca do ensino de física moderna e contemporânea:

\begin{citacao}
  Despertar a curiosidade dos estudantes e ajudá-los a reconhecer a Física como um empreendimento humano e, portanto, mais próxima a eles; os estudantes não têm contato com o excitante mundo da pesquisa atual em Física, pois não vêem nenhuma Física além de 1900. Esta situação é inaceitável em um século no qual idéias revolucionárias mudaram a ciência totalmente; é do maior interesse atrair jovens para a carreira científica. Serão eles os futuros pesquisadores e professores de Física \cite[p.~24]{ostermann}.
\end{citacao}

Além disso, a importância do ensino de Física não se resume ao conhecimento técnico \textit{per se}, mas também na alfabetização científica e no auxílio na formação de cidadãos com um olhar crítico aos problemas contemporâneos e que saibam atuar numa sociedade constantemente influenciada pelo progresso científico e pelos meios de comunicação digitais.

O objetivo deste estudo é apresentar aplicações diretas da espectroscopia astronômica como material de apoio para aulas de Física, tornando os tópicos de física moderna mais envolventes e acessíveis aos estudantes, ajudando-os a ver a relevância da Física no cotidiano. O estudo da espectroscopia permite aos estudantes a compreensão do surgimento e evolução do universo, conforme a habilidade EM13CNT201, assim como a representação e interpretação de modelos explicativos, dados e experimentos, o que se alinha com a habilidade EM13CNT301 da BNCC de Ciências da Natureza e suas Tecnologias. Além do mais, essas técnicas também envolvem os estudantes na interpretação de dados, gráficos e tabelas, consoante à habilidade EM13MAT102 da BNCC da disciplina de Matemática.

A metodologia utilizada para esta pesquisa se dá com a seleção dos tipos de corpos celestes mais reconhecidos: estrelas, nebulosas, galáxias (incluindo galáxias ativas), cometas, planetas e satélites naturais. A partir das medições de espectrometria de cada tipo de objeto celeste, obtidas através de pesquisa bibliográfica, analisaremos e apontaremos suas diferenças e características peculiares, a presença de linhas de emissão ou absorção em determinados comprimentos de onda, assim como a presença de possíveis desvios (\textit{redshift}) e suas magnitudes nas bandas vermelha e azul, e associá-las com os conceitos de física moderna vistos no ensino médio e ensino superior.

\section{Fundamentação teórica}

A história da Astronomia remonta a antiguidade. Sua instrumentação e visão matematizada foi promovida pelos gregos e, principalmente, pelos povos árabes \cite{rocha}, o que, posteriormente, levou à Revolução Copernicana e avanços realizados por Johannes Kepler, Galileu Galilei e Isaac Newton. No século XVII, Isaac Newton observou a luz do Sol passando por uma pequena fenda na janela, produzindo um facho estreito de luz. Ao posicionar um prisma de vidro nesse facho de luz, Newton observou que a luz branca do Sol se dividiu em muitas outras cores. Não apenas isso, ele também fez uso de uma lente e outro prisma para recombinar as cores, formando novamente a luz branca. Newton havia demonstrado empiricamente que a luz branca, na verdade, era uma composição de todas as outras cores. Esse espalhamento da luz branca em um arco-íris de cores visíveis foi chamado de \textit{espectro contínuo}.

Contudo, um passo importante para o nascimento da astrofísica foi o início da busca pelas razões físicas e químicas dos fenômenos que já eram conhecidos há séculos, como as diferenças na cor das estrelas ou o surgimento de supernovas. Essa ciência começou a emergir quando Joseph von Fraunhofer (1787–1826), inventor do espectroscópio, descobriu as linhas de absorção no espectro da luz solar ao analisá-la com este instrumento. Esse avanço levou a pesquisas subsequentes para identificar os espectros das estrelas, conforme descrito por \textcite{fraunhofer} ao analisar o espectro de Sirius, na constelação de Canis Majoris:

\begin{citacao}
  Também fiz experimentos com o mesmo aparelho [espectroscópio] com a luz de algumas estrelas fixas de primeira grandeza. Mas como a luz dessas estrelas é muitas vezes mais fraca que a de Vênus, o brilho da imagem colorida também é muitas vezes menor. No entanto, vi, sem engano, na imagem colorida da luz de Sirius, três listras largas, que parecem não ter nenhuma semelhança com as da luz do Sol; uma dessas listras é verde e duas são azuis. Listras também podem ser vistas na imagem colorida da luz de outras estrelas fixas de primeira magnitude; no entanto, essas estrelas, quanto às listras, parecem distintas entre si \cite[p.~220-221, tradução do autor]{fraunhofer}.
\end{citacao}

\begin{figure}[h!]
	\centering
	\caption{\textbf{Espectro de uma estrela classe A}}\label{sirius}
	\includegraphics[width=.6\textwidth]{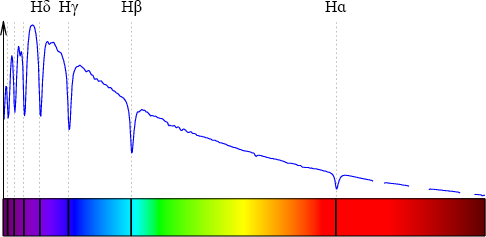}
	\source{Nicole Vogt/NMSU (adaptado pelo autor).}
\end{figure}

Hoje, sabe-se que as três listras largas observadas por Fraunhofer no espectro de Sirius são as fortes linhas de absorção do hidrogênio (Hβ, Hγ e Hδ) que, de acordo com \textcite{karttunen}, são características marcantes em estrelas brancas da classe A. Essas linhas podem ser observadas na Figura \ref{sirius} como grandes fendas nos comprimentos de onda do ciano ao violeta.

Em 1814, Joseph von Fraunhofer havia identificado e catalogado cerca de 475 linhas escuras no espectro da luz do Sol, estas que ficaram conhecidas como as \textit{linhas de Fraunhofer}. Poucos anos mais tarde, Gustav Kirchhoff determinou que 70 das linhas de Fraunhofer correspondiam a 70 linhas de vapor de ferro, o que marcou o nascimento da ciência da espectroscopia \cite{carroll}.

\begin{figure}[h!]
	\centering
	\caption{\textbf{As linhas de Fraunhofer do espectro solar}}\label{fraunhofer}
	\includegraphics[width=.9\textwidth]{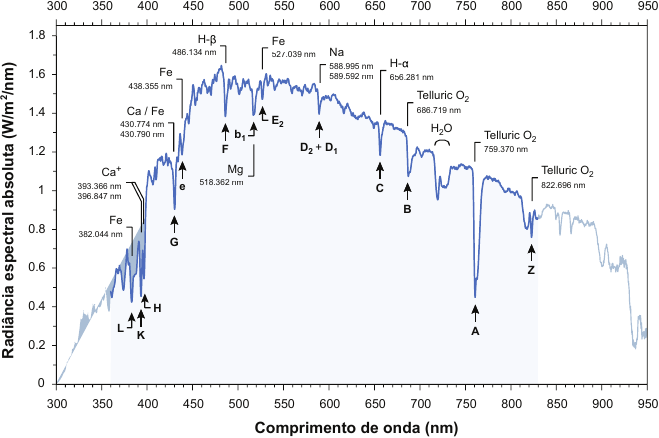}
	\source{Cyamahat/Wikimedia Commons (adaptado pelo autor). Disponível em: \url{https://commons.wikimedia.org/wiki/File:Solar_spectral_irradiance.svg}.}
\end{figure}

Em 1860, Gustav Kirchhoff e Robert Bunsen publicaram seus trabalhos, tendo Kirchhoff formulado de forma empírica suas três leis para a espectroscopia \cite{kepler,carroll}:

\begin{enumerate}
  \item Para o \textit{espectro contínuo}, um objeto sólido ou um gás opaco muito quente e denso emite radiação em todos os comprimentos de onda, implicando que a radiação de um corpo negro depende somente de sua temperatura, não de sua composição.
  \item Para o \textit{espectro de emissão}, um gás quente e pouco denso produz um espectro discreto de linhas brilhantes. A quantidade e o comprimento de onda dessas linhas dependem do elemento químico e do grau de ionização desse gás.
  \item Para o \textit{espectro de absorção}, se um espectro contínuo atravessar um gás em condições ideais de temperatura e pressão, esse gás absorve fótons nos mesmos comprimentos de onda de seu espectro de emissão, gerando as linhas escuras.
\end{enumerate}

Dá-se o nome de \textit{espectroscopia} ao estudo da decomposição da luz em seus comprimentos de onda ao atravessar um prisma ou uma rede de difração \cite{kepler}. A espectroscopia astronômica é aplicada por meio de espectrômetros desenvolvidos para a decomposição e medição de intensidade da luz de seus comprimentos de onda. Segundo \textcite{kitchin}, há duas maneiras de se fazer a espectroscopia astronômica: por interferência, que usa redes de difração ou interferômetros, ou por refração diferencial, que utiliza prismas.

\begin{figure}[h!]
  \centering
  \caption{\textbf{Plotagem do espectro de absorção e emissão do hidrogênio}}\label{emission_absorption}
  \includegraphics[width=.65\textwidth]{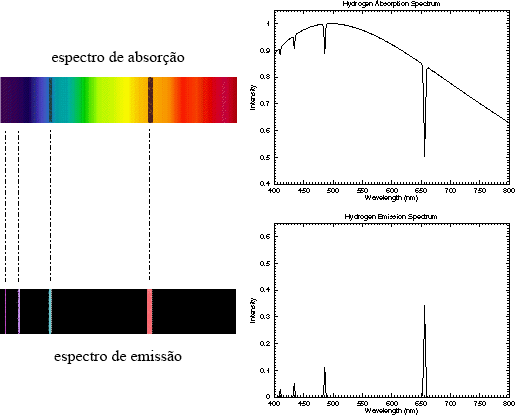}
  \source{STROBEL, 2022 (adaptado pelo autor). Disponível em: \url{https://www.astronomynotes.com/light/s5.htm}}
\end{figure}

A apresentação de dados obtidos por espectrômetros se faz com a plotagem de um gráfico de linha com os comprimentos de onda no eixo $x$ e a intensidade da emissão no eixo $y$. As linhas de emissão representam luz emitida em comprimentos de onda específicos por gás quente ou excitado, que irradia luz em comprimentos de onda específicos que indicam sua composição e são representadas como picos no gráfico. As linhas de absorção são comprimentos de onda onde a luz foi absorvida, removida do espectro observado por gás ou poeira que se encontra entre o detector e a fonte e, no gráfico, são identificadas como vales (Figura \ref{emission_absorption}).

Como descrevem \textcite{kepler}, no final do século XIX essas técnicas de observação foram aplicadas imediatamente em outras estrelas, levando à descoberta de um elemento ainda desconhecido, o hélio (He). No entanto, ainda faltava uma explicação plausível para o motivo da existência das linhas espectrais das estrelas. \textcite{carroll} apontam a extensão desse problema:

\begin{citacao}
  As linhas de absorção produzidas pelo hidrogênio são muito mais fortes para Vega do que para o Sol. Isso significa que a composição de Vega contém significativamente mais hidrogênio? A resposta é não. [...] A resposta exigia uma nova compreensão da natureza da própria luz \cite[p.~116, tradução do autor]{carroll}.
\end{citacao}

Hoje, sabe-se que as diferenças nas linhas espectrais se dão devido às diferenças de temperatura e pressão na fotosfera das estrelas. Estrelas muito quentes ionizam o hidrogênio, arrancando seus elétrons e impedindo a formação de linhas de absorção. Já as estrelas mais frias concentram a maior parte do hidrogênio no estado fundamental, também sem a possibilidade de produzir linhas espectrais \cite{kepler}.

Para \textcite{carroll}, a aplicação direta da Física na astronomia, a \textit{astrofísica}, obteve sucesso em explicar os espectros observados e contribuiu com o desenvolvimento de cálculos e teorias que levaram às descobertas de novos fenômenos como os buracos negros, os surtos de raios gama, as estrelas de nêutrons, entre muitos outros nunca imaginados.

Atualmente, quase todos os aspectos da astronomia estão intimamente ligados à Física. A física de partículas explica a existência e a interação dos raios cósmicos com nossa atmosfera \cite{karttunen}, a termodinâmica e a física nuclear explicam a estrutura e a dinâmica das estrelas em evolução, e a física da matéria condensada ainda visa explicar o misterioso interior de seus remanescentes, como as anãs brancas e estrelas de nêutrons \cite{kepler,horvath:estelar}.

\subsection{A luz como onda eletromagnética}

As equações de Maxwell definidas no século XIX descrevem os fenômenos elétricos e magnéticos como partes de uma mesma entidade, o eletromagnetismo. As equações também mostram que variações no campo elétrico e no campo magnético se propagam no espaço como ondas, chamadas de \textit{ondas eletromagnéticas} \cite{oliveira}. Diversas medições posteriores, como o experimento de Michelson-Morley, provaram que a luz era, também, uma onda eletromagnética.

A luz é uma das grandes portadoras da informação do universo. A radiação eletromagnética, juntamente com as ondas gravitacionais, os raios cósmicos e os neutrinos, formam os quatro mensageiros da astrofísica \cite{branchesi}. De acordo com \textcite{lena}, as características de emissão da radiação eletromagnética por um corpo celeste depende diretamente da natureza do emissor, incluindo seu movimento, temperatura, pressão, campos magnéticos, entre outras variáveis. Portanto, observar e analisar a luz emitida por um corpo nos permite determinar diversas propriedades físicas desse emissor.

Segundo \textcite{gallaway}, a \textit{fotometria}, ou seja, a medição da intensidade da radiação eletromagnética, quando realizada e analisada em um intervalo de tempo, também pode ser usada para estudar objetos como sistemas binários e variáveis cataclísmicas, discos de acreção, asteroides e estrelas variáveis Cefeidas, assim como a detecção de supernovas. Fontes intensas de radiação como estrelas de nêutrons e quasares cobrem todas as bandas do espectro eletromagnético, dos raios gama às ondas de rádio \cite{branchesi,horvath:altasenergias}, permitindo uma grande diversidade de análises por diferentes especialidades da Astrofísica.

\subsubsection{O espectro eletromagnético e a radiação de corpo negro}

Da mesma forma que o som pode ser dividido em frequências graves e agudas, infrasom e ultrassom, as ondas eletromagnéticas também são classificadas de acordo com sua frequência e comprimento de onda como diferentes regiões de um mesmo \textit{espectro eletromagnético}, como mostra a Tabela \ref{espectro} a seguir.

\begin{table}[h!]
	\centering
	\caption{\textbf{O espectro eletromagnético}}\label{espectro}
	\begin{tabular}{l l l}
		\hline
		\textbf{Banda} & \textbf{Frequência} & \textbf{Comprimento de onda} \\
		\hline
		Rádio & Abaixo de 300 MHz & > 1 m \\
		Micro-ondas & 300 MHz – 300 GHz & 1 m – 1 mm \\
		Infravermelho & 300 GHz – 430 THz & 1 mm – 700 nm \\
		Luz visível & 430 THz – 750 THz & 700 nm – 400 nm \\
		Ultravioleta & 750 THz – 30 PHz & 400 nm – 10 nm \\
		Raios X & 30 PHz – 30.000 PHz & 10 nm – 0,01 nm \\
		Raios gama & Acima de 30.000 PHz & < 0,01 nm \\
		\hline
	\end{tabular}
	\source{\textcite{karttunen}}
\end{table}

Um menor comprimento de onda de luz não significa somente uma maior frequência, mas também que esses fótons carregam mais energia, conforme a equação

\begin{equation}
  E = h \nu \equiv \frac{hc}{\lambda}
\end{equation}

\noindent{onde $h$ é a constante de Planck, $\nu$ é a frequência do fóton e $\lambda$ o comprimento de onda do fóton \cite{eisberg}.}

O comprimento de onda dos fótons emitidos por um objeto astronômico têm relação direta com suas propriedades físicas. Suponha um objeto hipotético que absorveria perfeitamente toda a radiação eletromagnética incidente, então, ele pareceria completamente preto, objeto que chamamos de \textit{corpo negro}. O fenômeno que muito intrigava os físicos até o início do século XX era que quando um corpo negro é aquecido, ele começa a emitir luz, a chamada \textit{radiação de corpo negro}. O fato é que ``todos os corpos negros à mesma temperatura emitem radiação térmica com o mesmo espectro'' \cite[p.~20]{eisberg}. Se esse corpo negro continuar a ser aquecido, os comprimentos de onda emitidos também mudam, brilhando em vermelho, depois em laranja, em amarelo e assim por diante, em comprimentos de onda cada vez mais curtos do espectro eletromagnético à medida que se aquece.

Dessa forma, tomemos como exemplo a relação entre temperatura e a radiância espectral definida por Kirchhoff em 1859. Gases extremamente quentes, na ordem dos milhões de Kelvin, têm emissões predominantes nos raios X, como os encontrados no disco de acreção de um buraco negro. No extremo oposto temos a radiação cósmica de fundo, extremamente fria, com temperatura em torno de 2,7 K, com comprimento de onda na faixa do micro-ondas \cite{silva:mclean}. Na seção \ref{high-low-energy} trataremos de outras formas de emissões no espectro de mais alta e mais baixa energia.

\begin{figure}[h!]
	\centering
	\caption{\textbf{Lei de deslocamento de Wien}}\label{wien}
	\includegraphics[width=.6\textwidth]{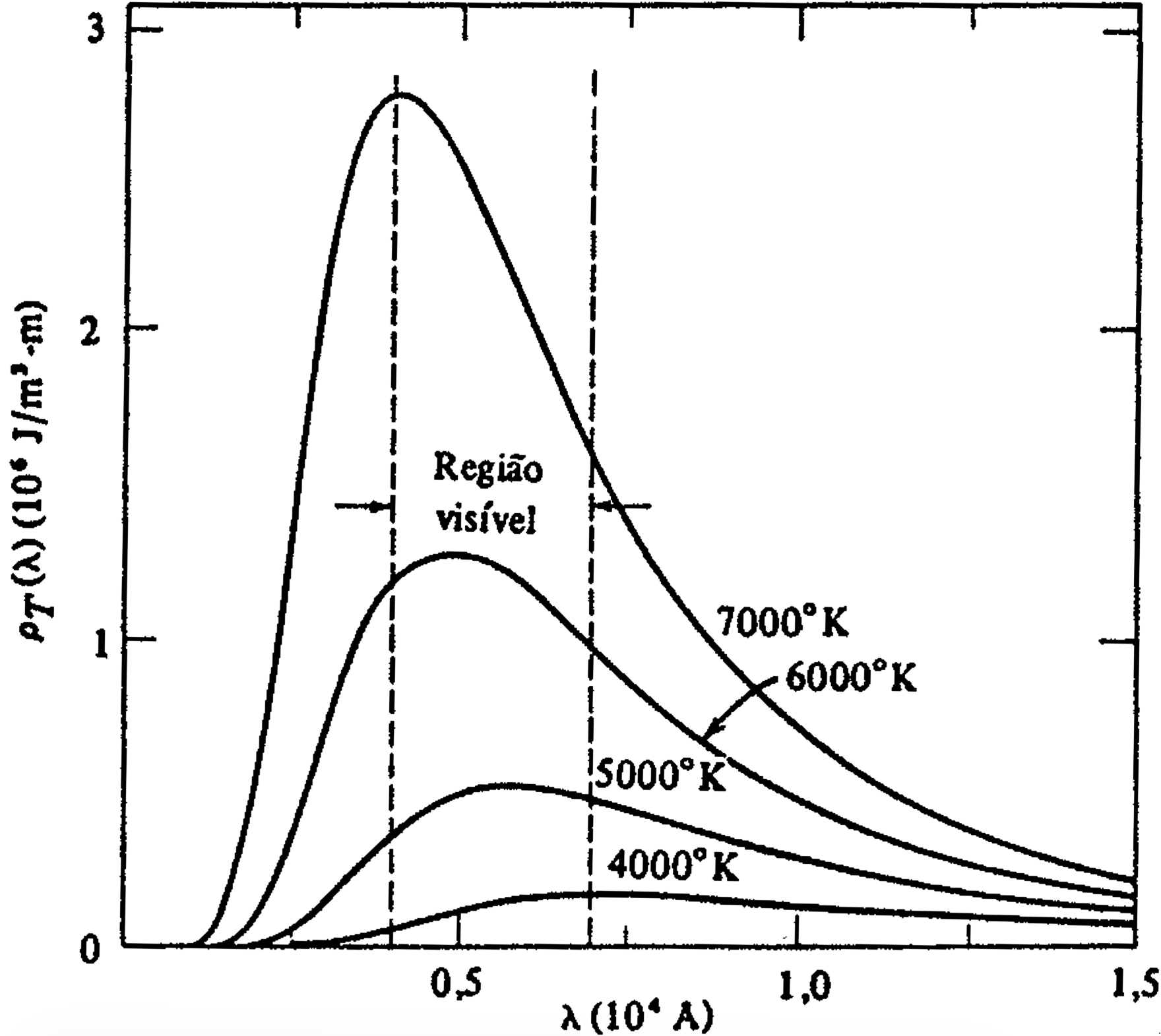}
	\source{\textcite{eisberg}}
\end{figure}

Estrelas são excelentes modelos de corpos negros pois emitem radiação, não absorvem, portanto, a cor aparente das estrelas é um indicativo direto de sua temperatura, onde o comprimento de onda dominante cresce de maneira inversamente proporcional à sua temperatura superficial, satisfazendo a condição da lei de deslocamento de Wien\footnote{Apesar de a lei de Wien e a lei de Rayleigh-Jeans fornecerem aproximações para o cálculo da radiação de corpo negro, esse fenômeno foi desvendado definitivamente por Max Planck, ao tratar a energia como uma variável discreta, ao invés de uma variável contínua.} (Figura \ref{wien}),

\begin{equation}
  \lambda_{\mathrm{max}} T = b
\end{equation}

\noindent{onde $b$ é a constante de dispersão, de valor aproximado de $2,9 \times 10^{-3}$ $\mathrm{m \cdot K}$, e $\lambda_{\mathrm{max}}$ o comprimento de onda no qual a radiância espectral atinge seu valor máximo para uma determinada temperatura $T$ em Kelvin \cite{horvath:estelar}, motivo pela qual Betelgeuse ($3.600 \pm 200$ K) é uma estrela vermelha, enquanto Bellatrix ($21.800 \pm 150$ K) tem coloração azulada. Quando $T$ aumenta, então $\lambda$ deve diminuir para que a constante $b$ seja preservada.}

Quando a temperatura de um corpo negro aumenta, este não apenas tem sua radiância máxima em comprimentos de onda mais curtos, mas também emite mais energia por unidade de área em todos os outros comprimentos de onda \cite{carroll}. Estudos realizados por Josef Stefan demonstraram que uma luminosidade $L$ aumenta em função da temperatura $T$ na quarta potência, conforme a equação de Stefan-Boltzmann

\begin{equation}
  L = A\sigma T^4
\end{equation}

\noindent{onde $A$ é a área do emissor e $\sigma$ a constante de Stefan-Boltzmann.}

A temperatura é apenas uma das variáveis que interferem nas medições de radiação eletromagnética. Podemos citar as lentes gravitacionais, capazes de distorcer o caminho da luz até o observador, e a extinção causada por nuvens de poeira localizadas entre o objeto e o observador \cite{lena}. Além disso, as emissões em longos comprimentos de onda, como a radiação cósmica de fundo em micro-ondas, também são um indicativo da expansão do universo. Conforme \textcite{silva:mclean}:

\begin{citacao}
  Como a velocidade da luz é finita e o Universo está se expandindo, o comprimento de onda de radiação EM [eletromagnética] emitida pelo objeto é esticado durante sua jornada para a Terra e se torna mais vermelho. Portanto, características espectrais críticas para sondar condições físicas e mecanismos em seu ponto de criação devem ser observadas em comprimentos de onda mais longos \cite[p.~3, tradução do autor]{silva:mclean}.
\end{citacao}

\subsection{O modelo de Bohr e as linhas espectrais}

Segundo \textcite{halliday}, no modelo atômico desenvolvido por Niels Bohr para o átomo de hidrogênio, a energia de um elétron não pode assumir valores arbitrários, portanto, esses elétrons devem ocupar níveis de energia definidos, com um momento angular múltiplo de $h/2\pi$, ou $\hbar$, de forma que

\begin{equation}
  m_e vr = n\hbar \equiv n\frac{h}{2\pi}
\end{equation}

\noindent{onde $m_e$ é a massa do elétron, $v$ sua velocidade, $r$ o raio da órbita do elétron, $h$ é a constante de Planck e $n$ o número quântico do nível de energia expresso como um número inteiro positivo (onde $n \in \mathrm{N^*}$).}

Um átomo pode absorver ou emitir fótons apenas com energia correspondente à diferença de energia de uma camada eletrônica a outra ($E = E_{n_2} - E_{n_1}$). Quando um fóton com energia suficiente para deslocar um elétron para um nível de energia mais elevado atravessa um gás, esse fóton é absorvido, transferindo sua energia para o elétron e fazendo-o se deslocar a um orbital superior. Quando esse elétron retorna ao seu estado original, essa energia extra é reemitida na forma de um fóton. Quanto maior a diferença de energia entre os níveis, maior a energia do fóton emitido \cite{eisberg}.

\begin{figure}[h!]
  \centering
  \caption{\textbf{Diagrama de transições para o átomo de hidrogênio}}\label{hydrogen}
  \includegraphics[width=.9\textwidth]{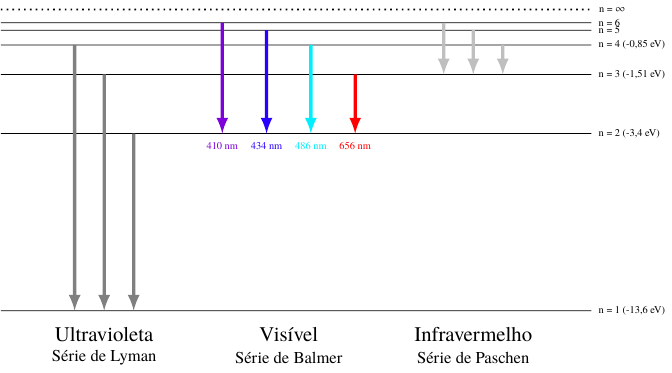}
  \source{Do autor.}
\end{figure}

Para um átomo de hidrogênio (Figura \ref{hydrogen}), isso significa que um elétron transitando da camada $n = 2$ ($-3,39$ eV) para a camada $n = 1$ ($-13,59$ eV), a diferença de energia de 10,20 eV corresponde à emissão de um fóton de comprimento de onda 121,6 nm, no espectro do ultravioleta, uma das emissões da série de Lyman. Um elétron transitando da camada $n = 3$ ($-1,51$ eV) para a camada $n = 2$ ($-3,39$ eV), a diferença de energia corresponde a um fóton de comprimento de onda 656,28 nm, de cor vermelha, uma linha espectral da série de Balmer chamada de hidrogênio-alfa (Hα).

Em termos de transições de uma camada $n_i$ para uma camada $n_f$ (onde $n_f > n_i$) em um átomo de hidrogênio, a função para se obter o comprimento de onda $\lambda$, em metros, de um fóton no vácuo pode ser expressa a partir da equação de Rydberg

\begin{equation}\label{rydberg}
  \frac{1}{\lambda} = R_H \left( \frac{1}{n_i^2} - \frac{1}{n_f^2} \right)
\end{equation}

\noindent{onde $R_H$ é a constante de Rydberg para o átomo de hidrogênio, de valor $1,097 \times 10^7$ $\mathrm{m^{-1}}$.}

Segundo \textcite{eisberg}, de forma generalizada, também pode-se usar o quarto postulado de Bohr

\begin{equation}
  \nu = \frac{E_i - E_f}{h}
\end{equation}

\noindent{onde a frequência de um fóton ($\nu$) pode ser determinada a partir da diferença das energias inicial e final dividida pela constante de Planck ($h$).}

\subsection{Emissões de alta e baixa energia}\label{high-low-energy}

A transição de elétrons entre orbitais atômicos é apenas uma das formas de emissão de ondas eletromagnéticas que podem ser detectadas na astrofísica. As transições eletrônicas são responsáveis pelas emissões nas bandas do infravermelho, visível e ultravioleta, mas não estão relacionadas às emissões de raios X ou ondas de rádio. Segundo \textcite{lena}, essas regiões mais extremas do espectro têm origem em outros fenômenos regidos pela mecânica quântica (Tabela \ref{tab:transitions}), que também permitem definir características físicas de nuvens moleculares, supernovas ou do meio interestelar.

\begin{table}[h!]
	\centering
	\caption{\textbf{Exemplos de diferentes transições discretas}}\label{tab:transitions}
	\begin{tabular}{l l l}
		\hline
		\textbf{Transição} & \textbf{Energia [eV]} & \textbf{Banda} \\
		\hline
		Estrutura hiperfina & $10^{-5}$ & Radiofrequência \\
		Interação spin-órbita & $10^{-5}$ & Radiofrequência \\
		Rotação molecular & $10^{-2}-10^{-4}$ & Micro-ondas e infravermelho \\
		Vibração-rotação molecular & $1-10^{-1}$ & Infravermelho \\
		Estrutura fina atômica & $1-10^{-3}$ & Infravermelho \\
		Transição eletrônica & $10^{-2}-10$ & Ultravioleta, visível e infravermelho \\
		Transição nuclear & $> 10^4$ & Raios X e raios gama \\
		Aniquilação elétron-pósitron & $\gtrsim 10^4$ & Raios gama \\
		\hline
	\end{tabular}
	\source{\textcite{lena}}
\end{table}

\section{A Espectroscopia Astronômica e a Física Moderna}

Como fora extensivamente abordado na fundamentação teórica deste estudo, a espectroscopia tem uma relação direta e profunda com conceitos cada vez mais avançados da física moderna. Temas como a quantização da energia, a radiação de corpo negro, o espectro de emissão e absorção e o modelo atômico de Bohr fazem parte do currículo das aulas de ciências exatas no ensino médio, mas são extremamente teóricos e matematizados, portanto, tópicos importantes para a compreensão do universo em que vivemos e para o exercício da análise e interpretação de dados científicos (conforme a BNCC de Ciências da Natureza e de Matemática) se tornam distantes dos estudantes e de aplicações reais.

Apesar de a astronomia ser associada às observações no óptico, muitos dos avanços se deram aos estudos dos espectros dos corpos celestes, que representam suas propriedades, como composição, temperatura, densidade e movimento. Essas propriedades, associadas com as observações extremamente distantes (como as realizadas pelo telescópio James Webb) permitem a elaboração de modelos cada vez mais precisos sobre a composição química, a origem e evolução do universo — temas também associados à BNCC de Ciências da Natureza.

A seguir, serão apresentadas e analisadas as espectrometrias de diferentes corpos celestes, apontando suas características peculiares e o que representam, levando aos estudantes uma maior proximidade com a análise de dados e o pensamento científico ao compreender como a ciência da espectroscopia realmente explica a natureza do universo que nos rodeia.

\subsection{A importância da espectroscopia astronômica}

Quando usamos telescópios e instrumentos de detecção para obter dados astronômicos de um corpo celeste, sua forma, cor e brilho podem ser usados para determinar seu tipo, por exemplo, diferenciar uma estrela da sequência principal de uma anã branca, assim como suas características físicas gerais. À medida que buscamos observar objetos cada vez mais distantes, eles aparecem cada vez menores e, também por limitação da resolução dos telescópios, torna-se cada vez mais difícil determinar suas propriedades físicas apenas a partir de uma imagem. No entanto, os diversos tipos de objetos celestes, como estrelas, galáxias e quasares, podem ser facilmente diferenciados pelas características únicas de seus espectros.

\begin{figure}[h!]
  \centering
  \caption{\textbf{Espectro da luz solar}}\label{sol}
  \includegraphics[width=1\textwidth]{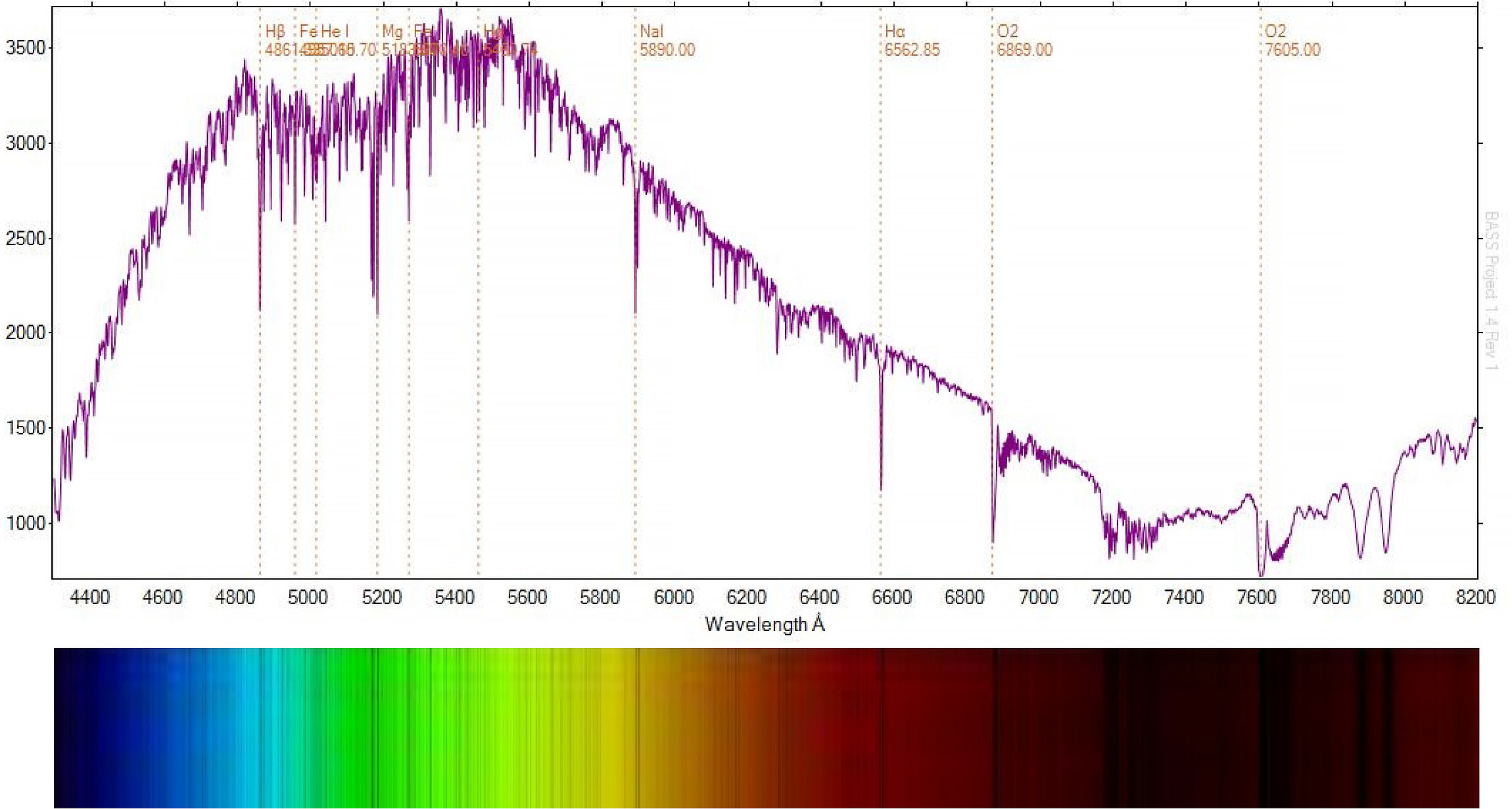}
  \source{UnderOak Observatory}
\end{figure}

Se a radiação de uma estrela tivesse sua fonte totalmente proveniente da energia térmica, seu espectro seria apresentado como uma curva suave como a de um corpo negro perfeito descrito pela lei de Planck. No entanto, quando observamos o espectro de uma estrela, o que vemos é uma série de vales correspondentes à absorção dos elementos químicos presentes na atmosfera dessa estrela, como mostra a Figura \ref{sol}, do espectro do Sol.

\begin{figure}[h!]
  \caption{\textbf{Objetos astronômicos luminosos em imagem no espectro visível}}\label{example_visible}
	\begin{subfigure}{0.5\textwidth}
		\centering
		\includegraphics[width=0.65\linewidth]{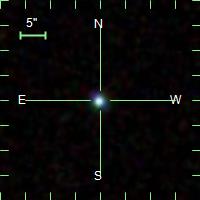}
		\caption{Objeto SDSS J005328.64-004321.3}\label{example_visible_star}
	\end{subfigure}
	\begin{subfigure}{0.5\textwidth}
		\centering
		\includegraphics[width=0.65\linewidth]{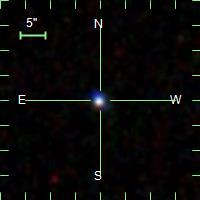}
		\caption{Objeto SDSS J005303.44-005436.8}\label{example_visible_quasar}
	\end{subfigure}
  	\source{\textcite{sdss}}
\end{figure}

Partindo para um exemplo prático, os dois objetos luminosos mostrados na Figura \ref{example_visible} em dados públicos disponibilizados pelo Sloan Digital Sky Survey (SDSS) parecem quase idênticos quando observados em um telescópio óptico convencional.

\begin{figure}[h!]
  \caption{\textbf{Espectrometria dos objetos astronômicos das Figuras \ref{example_visible_star} e \ref{example_visible_quasar}}}\label{example_spectra}
	\begin{subfigure}{0.5\textwidth}
		\centering
		\includegraphics[width=.95\linewidth]{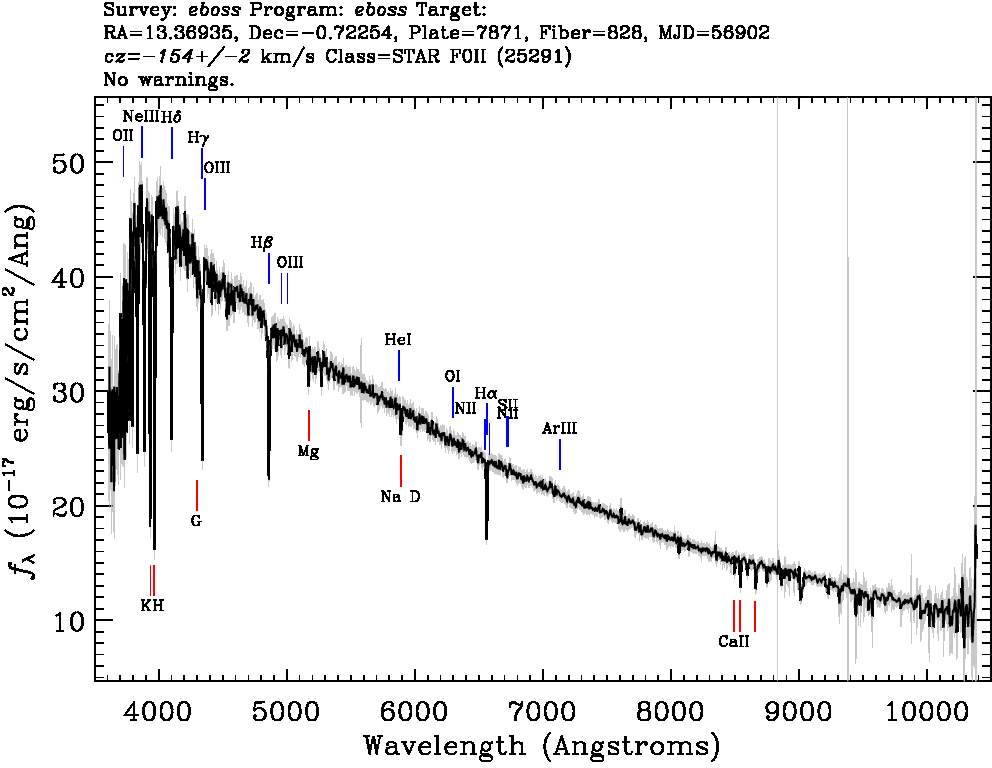}
  		\caption{Espectro da estrela da Figura \ref{example_visible_star}.}
	\end{subfigure}
	\begin{subfigure}{0.5\textwidth}
		\centering
		\includegraphics[width=.95\linewidth]{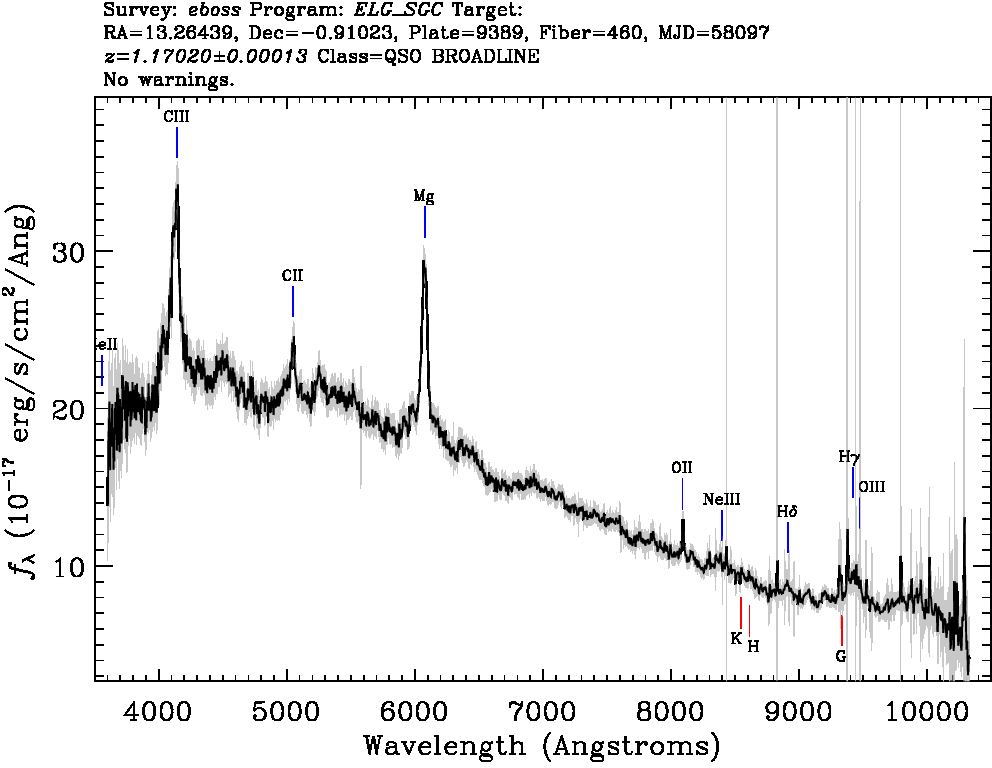}
		\caption{Espectro do quasar da Figura \ref{example_visible_quasar}.}
	\end{subfigure}
	\source{\textcite{sdss}}
\end{figure}

No entanto, o objeto da Figura \ref{example_visible_star} é uma estrela variável da classe F0II localizada em nossa própria galáxia, enquanto o objeto da Figura \ref{example_visible_quasar} é um quasar localizado a 26 bilhões de anos-luz de distância. Somente ao obtermos seus respectivos dados espectrais é possível notar que são objetos completamente distintos. Essa diferença pode ser vista na Figura \ref{example_spectra} a seguir.

Os dois objetos parecem quase idênticos em intensidade de brilho e cor quando em uma imagem em luz visível, portanto, sem a espectrometria desses objetos seria quase impossível dizer quais objetos nas imagens são estrelas e quais são quasares.

Medir o espectro de um objeto também nos permite determinar a distância de um objeto da Terra e medir outras propriedades físicas de interesse, como sua massa, idade, composição e velocidade radial em relação à Terra. Objetos se afastando da Terra têm seus comprimentos de onda ``esticados'' devido ao desvio Doppler da luz, portanto, detectar pequenos desvios nos picos e vales da espectrometria permite determinar a velocidade de um objeto se afastando ou se aproximando de nós, por meio da equação:

\begin{equation}
  z = \frac{\lambda_o - \lambda_e}{\lambda_e}
\end{equation}

Supondo o comprimento de onda emitido ($\lambda_e$) do hidrogênio-alfa de 656 nm, se analisarmos uma estrela cujo comprimento observado ($\lambda_o$) se encontra em 654 nm, obtemos

\begin{equation}
  z = \frac{654 - 656}{656} \approx -0,003
\end{equation}

\noindent{um desvio negativo, ou seja, essa estrela está se aproximando da Terra. Multiplicar esse valor pela velocidade da luz ($c$) nos permite encontrar a velocidade com que esse objeto se aproxima ou se afasta de nós — no caso do exemplo, $-0,003c \approx -899$ km/s.}

\subsection{Estrelas}

A superfície quente e densa das estrelas (fotosfera) emite radiação no espectro contínuo que, representada num gráfico, forma uma curva que se aproxima de um corpo negro. ``Átomos na atmosfera acima da superfície absorvem certos comprimentos de onda característicos dessa radiação'' \cite[p.~227]{karttunen}, logo, o que vemos nos dados obtidos de uma estrela é um espectro contínuo emitido pela fotosfera opaca contendo diversas lacunas que representam as linhas de absorção dos elementos químicos dos gases presentes na atmosfera.

\begin{table}[h!]
	\centering
	\caption{\textbf{A classificação espectral de Harvard}}\label{tab:harvard}
	\begin{tabular}{l l l l}
		\hline
		\textbf{Classe} & \textbf{Temperatura} & \textbf{Descrição da cor} & \textbf{Exemplo} \\
		\hline
		O & > 30.000 K & Azul & Alnitak \\
		B & 10.000 – 30.000 K & Branco-azulada & Rigel \\
		A & 7.500 – 10.000 K & Branca & Vega \\
		F & 6.000 – 7.500 K & Branco-amarelada & Procyon \\
		G & 5.000 – 6.000 K & Amarela & Sol \\
		K & 3.500 – 5.000 K & Laranja & Arcturus \\
		M & < 3.500 K & Vermelha & Betelgeuse \\
		\hline
	\end{tabular}
	\source{\textcite{leblanc}}
\end{table}

As estrelas são classificadas de acordo com sua temperatura superficial, a \textit{classificação espectral de Harvard} (Tabela \ref{tab:harvard}). Segue-se a ordem, da mais quente para a mais fria: O, B, A, F, G, K e M. As letras são fora de ordem por razões históricas, com as estrelas inicialmente ordenadas por intensidade das linhas de hidrogênio, mas, posteriormente, reordenadas por temperatura.

\begin{figure}[h!]
  \centering
  \caption{\textbf{Espectro visível de acordo com a classificação estelar}}\label{star_spectra}
  \includegraphics[width=1\textwidth]{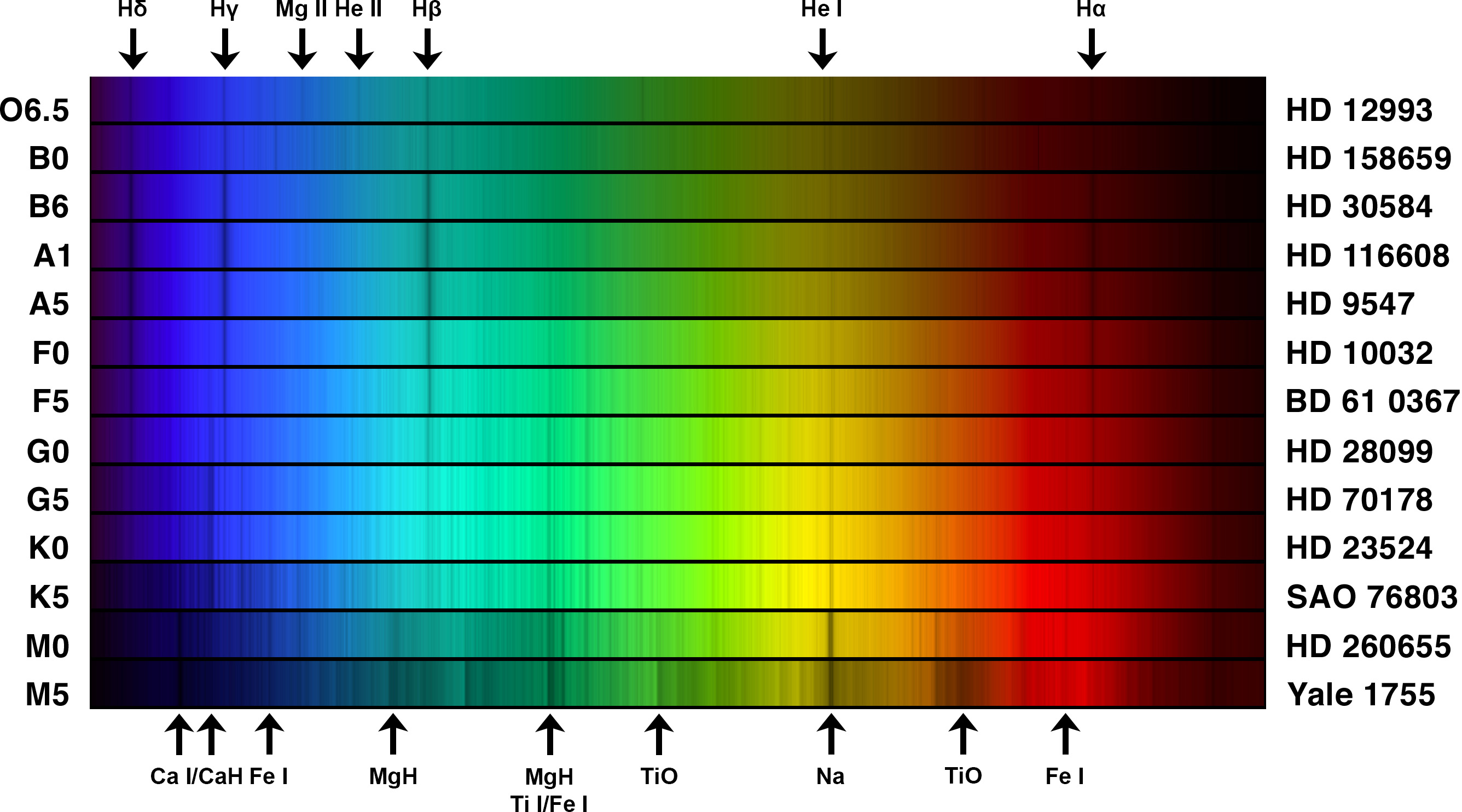}
  \source{NOIRLab/NSF/AURA (adaptado pelo autor). Disponível em: \url{https://noirlab.edu/public/images/noao0134a}.}
\end{figure}

Agora, seguiremos sobre o que os espectros podem nos revelar. Observe a Figura \ref{star_spectra} a seguir, com a dispersão do espectro visível de estrelas de classes distintas.

O primeiro fato observado é que estrelas quentes emitem luz mais intensa em comprimentos de onda mais curtos, próximos ao violeta, enquanto estrelas mais frias emitem luz em comprimentos de onda longos, se aproximando do vermelho. Essa é uma evidência clara da relação entre a lei de Wien para corpos negros, a temperatura e a cor de uma estrela.

\begin{figure}[h!]
  \centering
  \caption{\textbf{Relação entre temperatura e intensidade de linhas espectrais}}\label{temperature_spectra}
  \includegraphics[width=.7\textwidth]{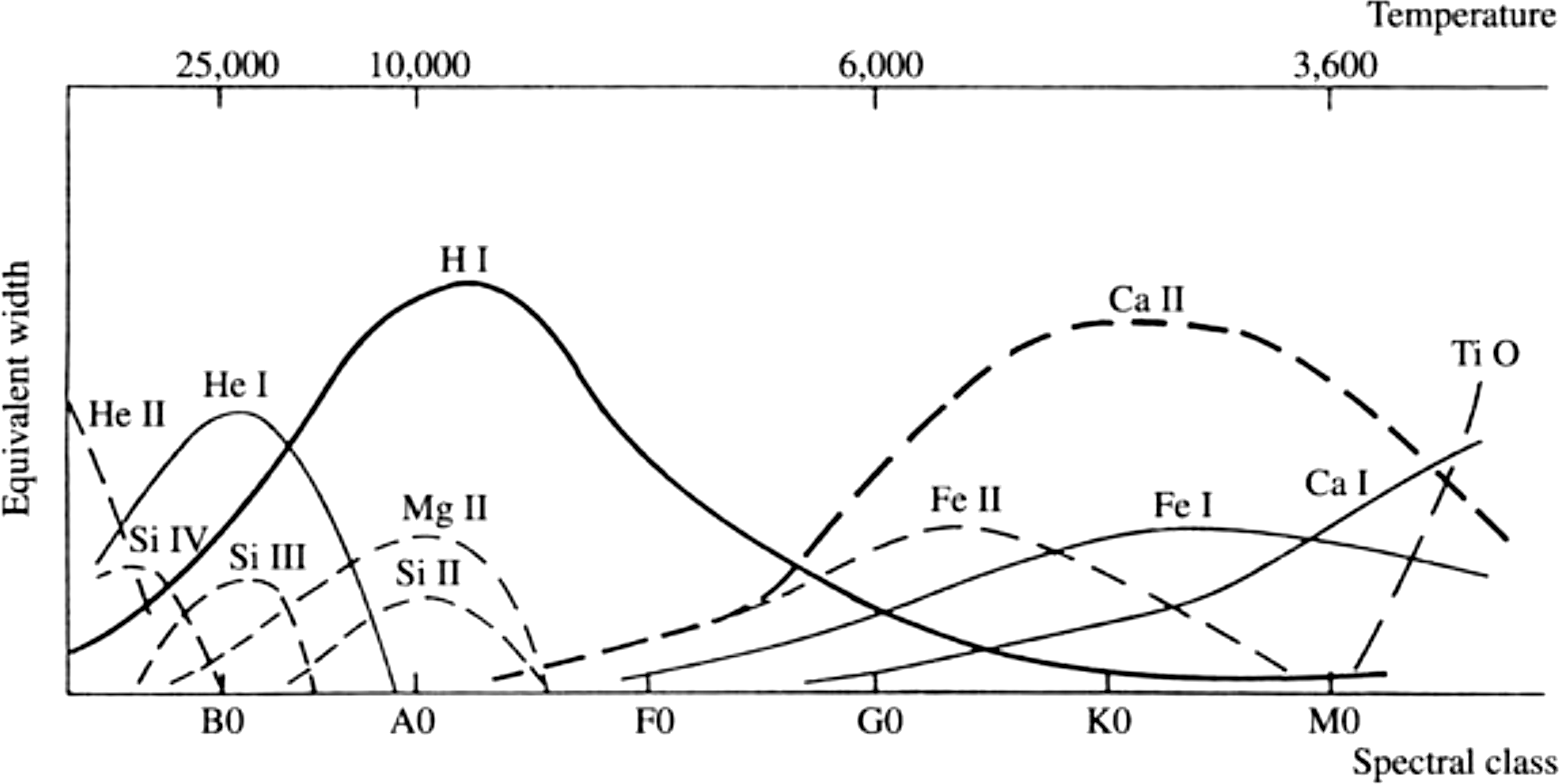}
  \source{\textcite{karttunen}}
\end{figure}

Os espectros das estrelas mais quentes, de classes O, B e A, diferem das mais frias, de classes K e M, não somente no que diz respeito ao pico da curva de emissões para um corpo negro, mas também em diferenças nas linhas de absorção dos elementos, sobretudo nas linhas de hidrogênio (Hα, Hβ, Hγ e Hδ) e sódio (Na). Isso, de fato, não significa que uma estrela possua maior quantidade de hidrogênio do que a outra. Como já fora abordado, uma estrela com uma temperatura superficial extremamente quente ou extremamente fria é incapaz de produzir linhas de absorção do hidrogênio, uma vez que seus átomos estão ionizados ou permanecem no estado fundamental, respectivamente \cite{karttunen}. Em uma temperatura ideal de aproximadamente 9.000 K, como de uma estrela classe A, grande parte dos átomos de hidrogênio têm seus elétrons excitados, produzindo as fortes linhas de absorção observadas em estrelas de classe B e A. O mesmo acontece para outros elementos químicos em seus diversos graus de ionização, conforme ilustra a Figura \ref{temperature_spectra}.

Estrelas frias da classe M, por outro lado, exibem fortes linhas de absorção de metais como ferro, sódio e cálcio. A baixa temperatura da superfície de uma estrela de classe M favorece a formação de moléculas, como o óxido de titânio (TiO) e óxido de vanádio (VO), que, em estrelas mais quentes, seriam destruídas por fotodissociação ou colisão \cite{leblanc}. Além disso, a estrutura do interior dessas estrelas é diferente de uma estrela como o Sol. As anãs vermelhas têm fortes correntes internas de convecção que ``misturam'' toda a sua composição interna, fazendo com que elementos metálicos presentes no interior emerjam na sua superfície em um fenômeno conhecido como \textit{dragagem} \cite{horvath:estelar}.

\subsubsection{Descontinuidade de Balmer}

\begin{figure}[h!]
  \centering
  \caption{\textbf{Linhas de Balmer para o átomo de hidrogênio}}\label{balmer-h}
  \includegraphics[width=.75\textwidth]{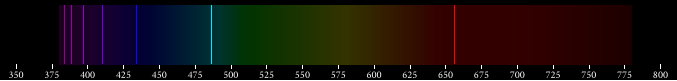}
  \source{Do autor.}
\end{figure}

O espectro das estrelas quentes também nos permite observar um fenômeno curioso. Repara-se que as linhas de Balmer para o hidrogênio se tornam cada vez mais próximas à medida que o comprimento de onda diminui, fato observável na Figura  \ref{balmer-h} abaixo.

\begin{figure}[h!]
	\centering
	\caption{\textbf{Descontinuidade de Balmer no espectro de Sirius}}\label{balmer-jump}
	\includegraphics[width=.55\textwidth]{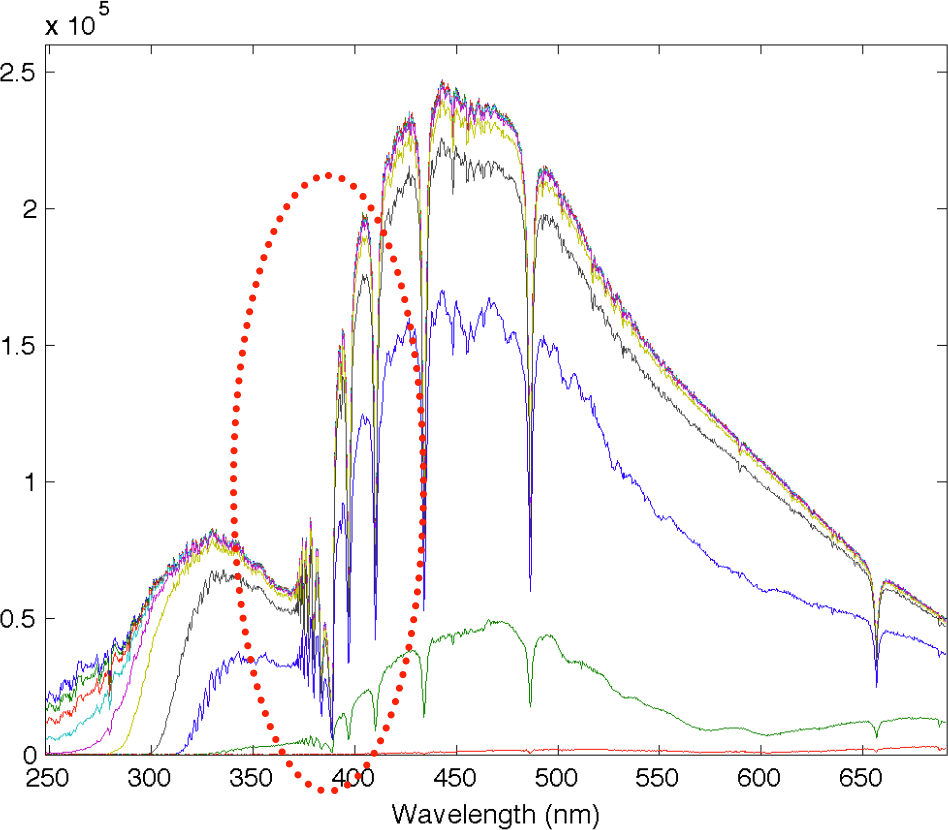}
	\source{\textcite{kyrola} (adaptado pelo autor).}
\end{figure}

De fato, estabelecer o limite $n_f \to \infty$ para a equação de Rydberg (eq. \ref{rydberg}) com $n_i = 2$ demonstra que as linhas de Balmer convergem para 364,5 nm, no espectro ultravioleta. Isso se deve ao fato de as camadas eletrônicas mais elevadas terem níveis de energia com diferença decrescente. Essa queda abrupta nas emissões causadas por linhas de absorção sobrepostas é a  \textit{descontinuidade de Balmer} e pode ser vista mais claramente na espectrometria de estrelas de classe B e A (Figura \ref{balmer-jump}).

\subsubsection{Estrelas tipo Wolf-Rayet}

O interior de uma estrela em seus estágios avançados de evolução é dividido em diversas camadas concêntricas, ou ``cascas'', compostas por elementos e temperaturas diferentes. De acordo com \textcite{horvath:altasenergias}, elementos leves, como hidrogênio e hélio que não sofreram fusão, se acumulam nas cascas superiores. Elementos mais pesados, como nitrogênio, carbono e oxigênio, são mais abundantes nas cascas inferiores, e assim por diante, até elementos extremamente densos como o silício, cálcio e ferro, que se são abundantes em seu núcleo.

\begin{figure}[h!]
  \centering
  \caption{\textbf{Dados espectrais de WR 140}}\label{wr140}
  \includegraphics[width=.8\textwidth]{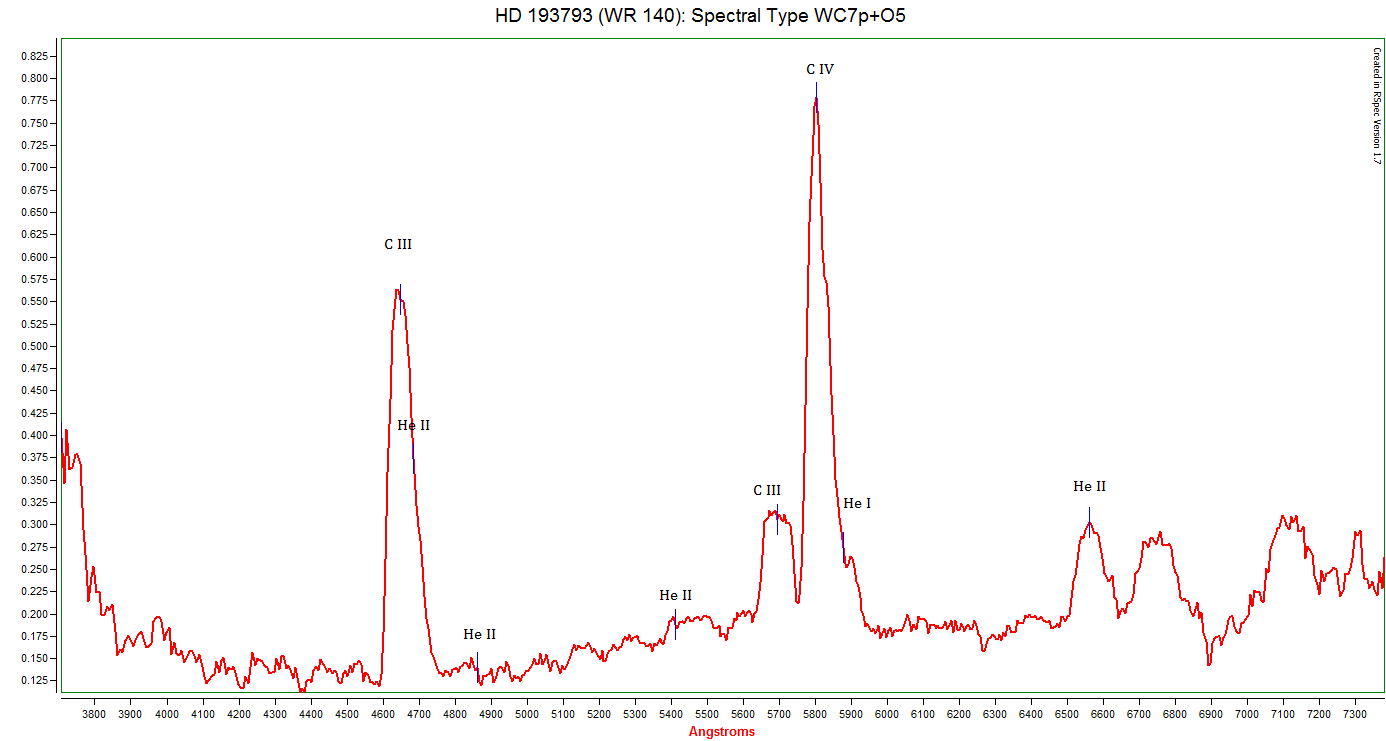}
  \source{UnderOak Observatory. Disponível em: \url{http://www.underoakobservatory.com/HighResImage_HD193793.html}}
\end{figure}

As estrelas do tipo Wolf-Rayet (também chamadas de estrelas W-R) são estrelas massivas, com mais de 30 massas solares e temperaturas acima de 30.000 K, que perderam suas cascas superiores de hidrogênio, deixando à mostra as cascas inferiores de hélio e carbono  \cite{horvath:estelar}. Isso significa que o espectro dessas estrelas é ausente de hidrogênio, apresentando uma atmosfera com grande variedade de linhas de emissão de hélio e carbono ionizados que podem ser vistas na Figura \ref{wr140}.

As altas temperaturas superficiais de uma estrela tipo Wolf-Rayet faz com que o carbono seja duplamente ou triplamente ionizado, indicado na espectrometria com os símbolos CIII e CIV. Existem, ainda, estrelas W-R que perderam seu envelope de hélio, isso faz com que o espectro dessas estrelas seja ausente tanto de hidrogênio quanto de hélio.

\subsection{Regiões H II e nebulosas planetárias}

\begin{figure}[h!]
  \caption{\textbf{Tipos de nebulosas de emissão}}\label{nebula_types}
	\begin{subfigure}{0.5\textwidth}
		\centering
		\includegraphics[width=1\linewidth]{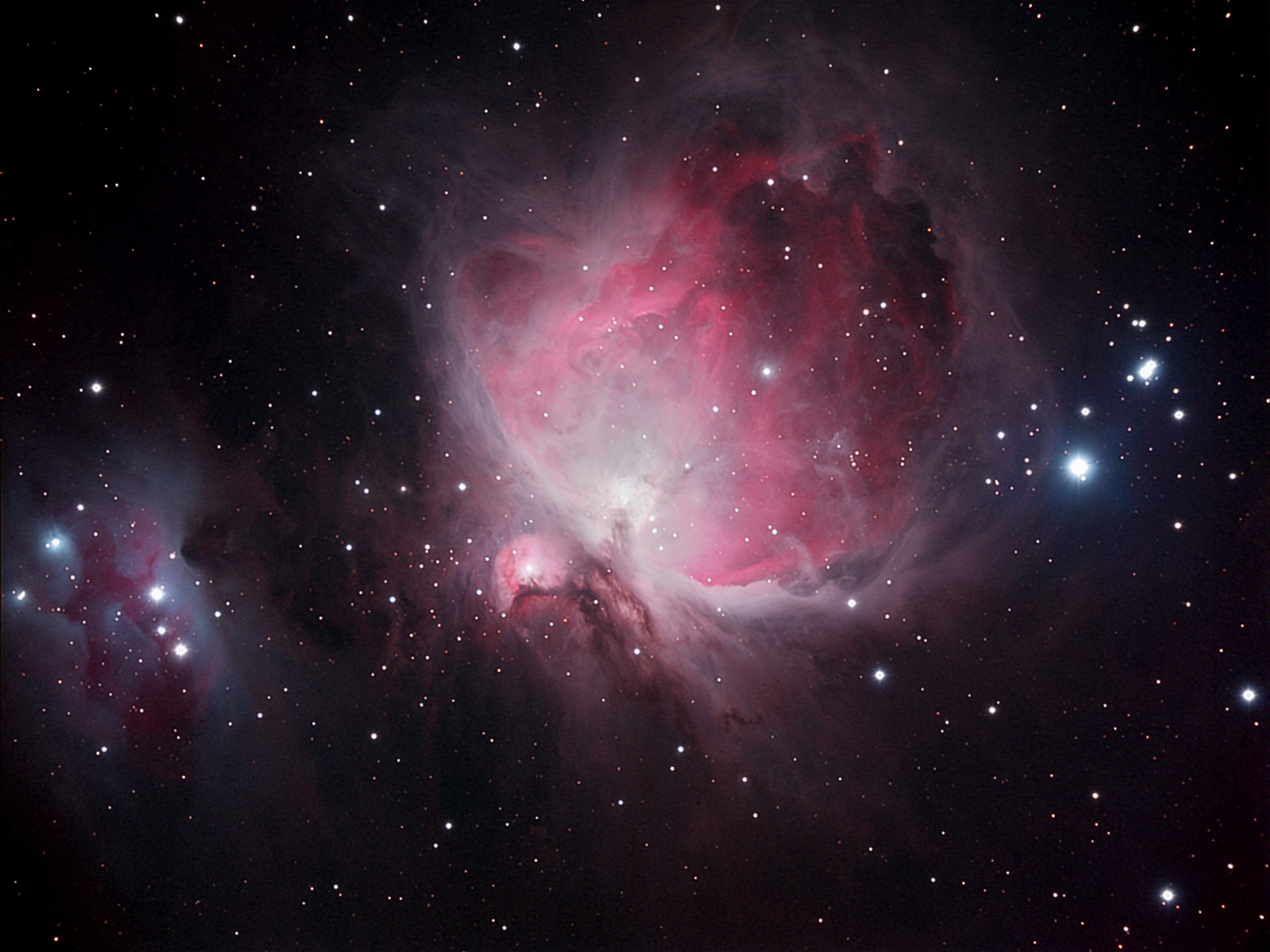}
		\caption{Nebulosa de Órion, uma região H II.}
	\end{subfigure}
	\begin{subfigure}{0.5\textwidth}
		\centering
		\includegraphics[width=1\linewidth]{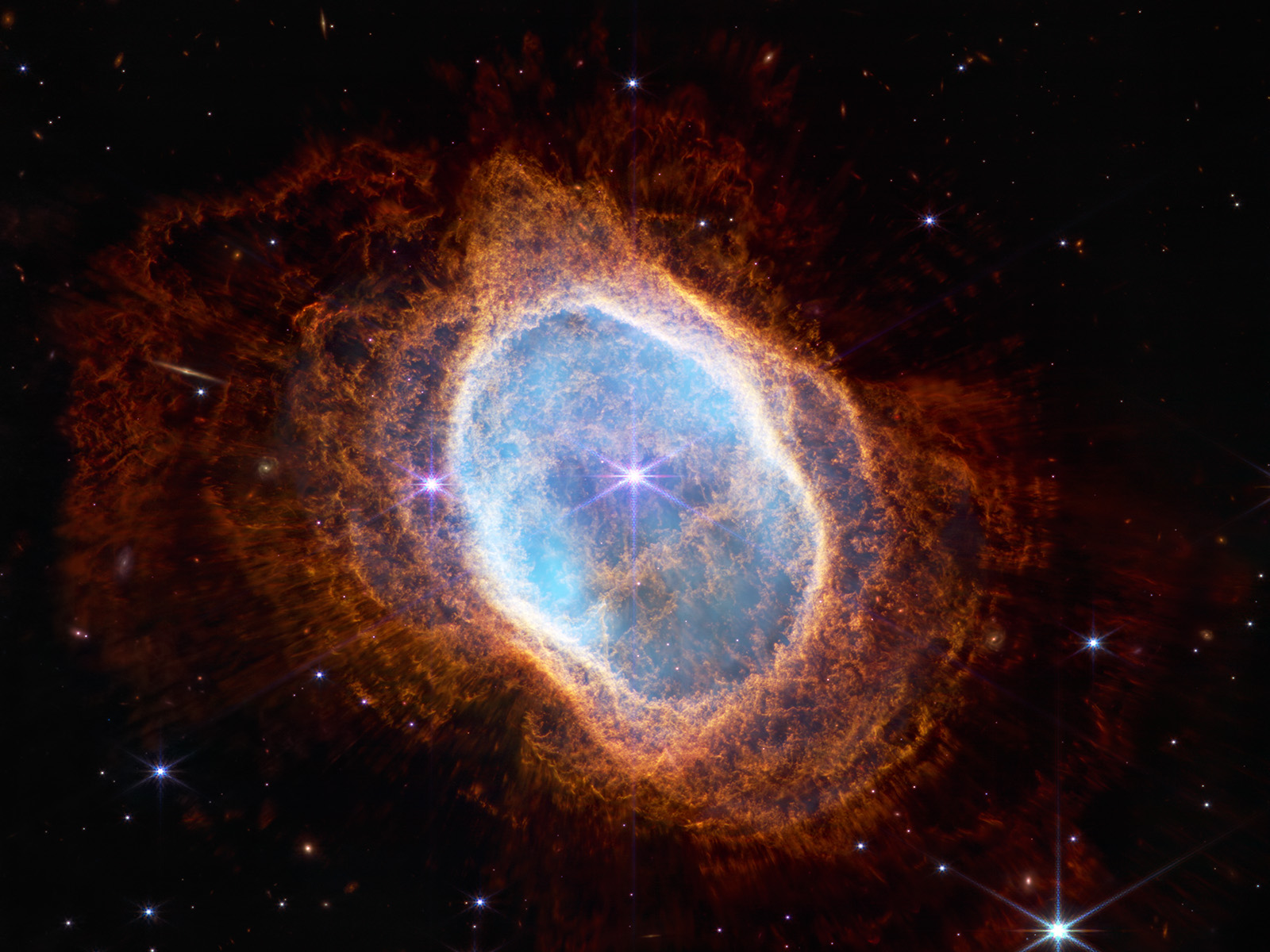}
		\caption{NGC 3132, uma nebulosa planetária.}
	\end{subfigure}
	\source{(a) Brian Davis. (b) NASA/ESA/CSA/STScI.}
\end{figure}

As nebulosas podem se dividir em duas grandes classificações: \textit{nebulosas escuras}, que não emitem luz e podem ser vistas como enormes regiões escuras no espaço, e as \textit{nebulosas de emissão}, que emitem luz própria \cite{kepler}. As regiões H II e as nebulosas planetárias são exemplos de nebulosas de emissão (Figura \ref{nebula_types}).

As regiões H II, também chamadas de ``berçários de estrelas'', são enormes nuvens de gás e poeira remanescentes de antigas supernovas que estão em fase de formação ativa de novas estrelas \cite{peimbert}. Essas estrelas jovens emitem radiação ultravioleta que ionizam e excitam o gás em seus entornos, fazendo-o brilhar. O exemplo de região H II mais conhecido é M42, a Nebulosa de Órion.

Uma nebulosa planetária é um remanescente de estrela de massa intermediária, tal como o Sol, e são bem menores do que as regiões H II. O que vemos em uma nebulosa planetária é o núcleo, uma anã branca, envolto por uma nuvem de gás ionizado que já fez parte da estrela genitora. Nossa estrela se transformará em uma nebulosa planetária, pois o Sol não tem massa suficiente para se colapsar e explodir em uma supernova, portanto, ao fim da vida, as fusões cessarão e os ventos solares ejetarão as camadas de gás, expondo o núcleo sólido degenerado.

Tanto as nebulosas planetárias quanto as regiões H II emitem luz própria. Por serem regiões de gás rarefeito, isso significa que seus espectros são dominados por linhas de emissão. Segundo \textcite{peimbert}, as estrelas que originaram as nebulosas planetárias eram estrelas de massa intermediária sem as condições ideais para a fusão do carbono, oxigênio e nitrogênio, portanto, seus espectros são dominados por esses tipos de elementos.

\begin{figure}[h!]
  \centering
  \caption{\textbf{Nebulosa do Anel (M57) vista por uma rede de difração}}\label{m57}
  \includegraphics[width=.75\textwidth]{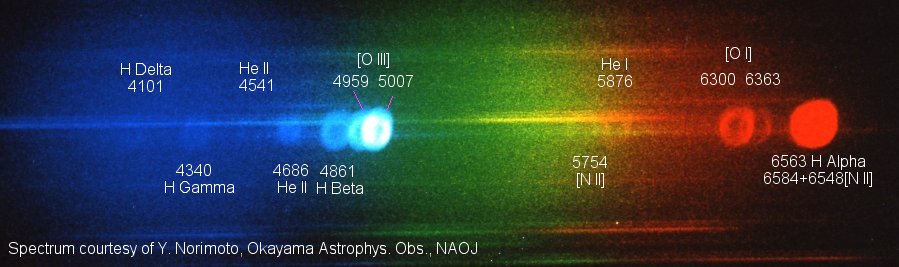}
  \source{Norimoto, Y./Okayama Observatory. Disponível em: \url{http://stars.astro.illinois.edu/sow/ring-p.html}.}
\end{figure}

A Figura \ref{m57} mostra uma nebulosa planetária quando vista por uma rede de difração. Observe que, ao contrário de uma estrela, que emite um espectro contínuo com lacunas de linhas de absorção, o espectro de uma nebulosa é inteiramente formado por emissão. Isso se deve ao fato de seu brilho se originar das emissões dos gases excitados que a compõem, de forma similar à uma lâmpada fluorescente. As linhas de emissão das nebulosas indicam a presença de gases das antigas estrelas que a originaram, como hidrogênio, hélio, oxigênio, nitrogênio e enxofre.

\subsubsection{Linhas proibidas}

A atmosfera da Terra ao nível do mar tem uma pressão de aproximadamente 101,3 kPa, isso faz com que os átomos e moléculas estejam em constante interação. No entanto, uma nebulosa é extremamente rarefeita, com densidade menor do que qualquer vácuo que possamos produzir em laboratório, dessa forma, as linhas espectrais em nebulosas apresentam algumas características peculiares.

As linhas espectrais são geradas muito rapidamente, com ``o tempo de vida dos elétrons em um estado excitado sendo de apenas $10^{-8}$ segundos'' \cite[p.~330]{bless}. Uma linha proibida é produzida quando um elétron permanece estável por mais tempo em um nível de energia superior por vários segundos, ou até horas, e salta espontaneamente para um nível inferior. Tal transição é dita ter uma probabilidade muito baixa, pois, em um gás de densidade maior, o átomo excitado colidiria com outros átomos ou elétrons livres e perderia energia muito antes de poder irradiar os fótons. No entanto, nas baixas densidades do meio interestelar e das nebulosas, entre 100 e 10.000 átomos/$\mathrm{cm^3}$, as interações são extremamente raras e há tempo para ocorrer o decaimento espontâneo desses orbitais chamados \textit{metaestáveis}.

\begin{figure}[h!]
  \centering
  \caption{\textbf{Espectrometria da nebulosa planetária PN Sextans A}}\label{pn_spectra}
  \includegraphics[width=.7\textwidth]{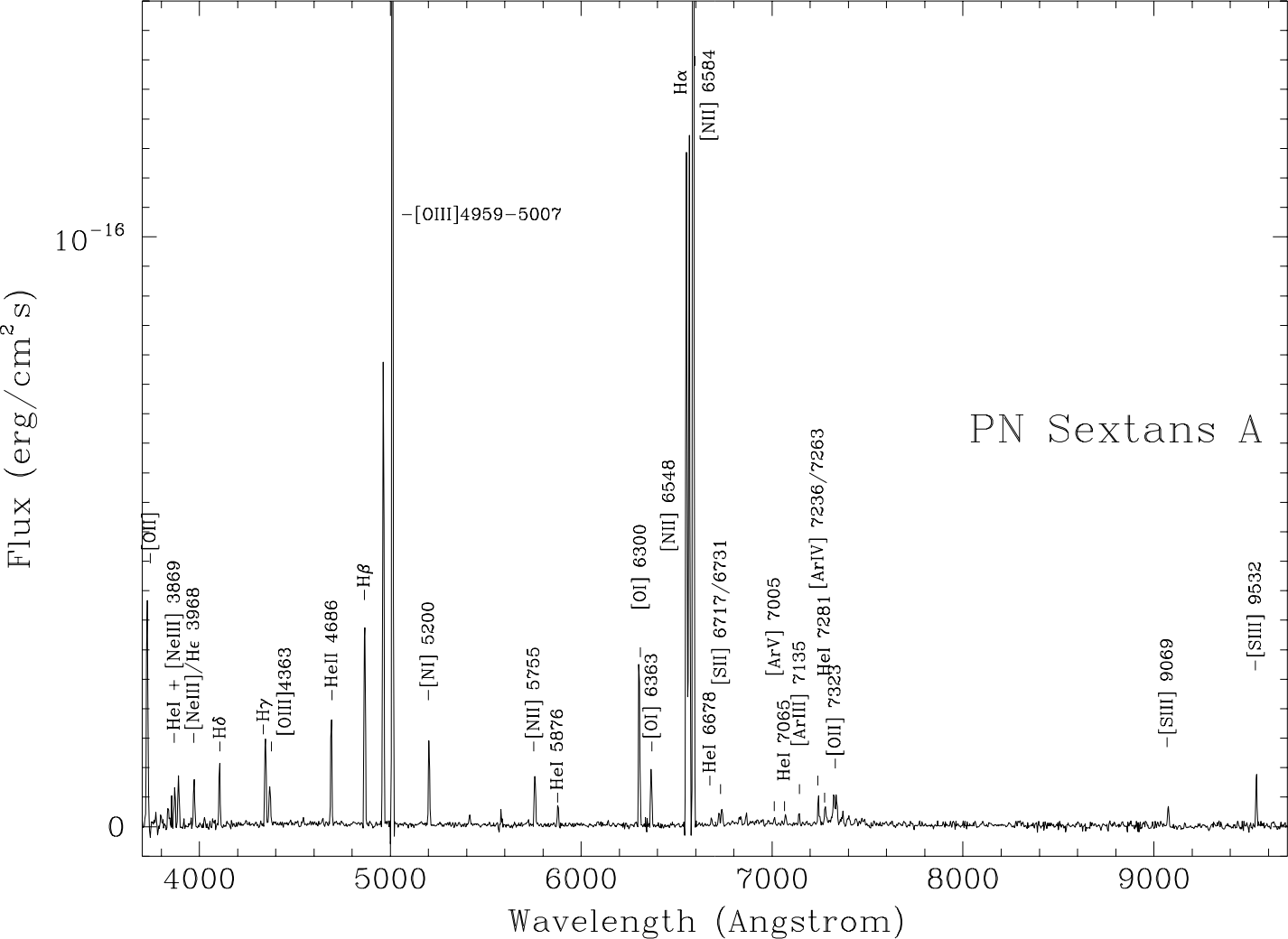}
  \source{\textcite{magrini}}
\end{figure}

As linhas proibidas são indicadas por colchetes e, como constatado, não são realmente proibidas, apenas têm uma chance muito baixa de acontecer. A Figura \ref{pn_spectra} apresenta dados espectroscópicos da nebulosa planetária PN Sextans A, mostrando a predominância de linhas proibidas de oxigênio duplamente ionizado ([OIII]) e de nitrogênio ionizado ([NII]).

As linhas proibidas desaparecem acima de uma certa densidade crítica (normalmente cerca de 108 átomos/$\mathrm{cm^3}$), sendo, portanto, sua existência um indicador de densidade do gás interestelar. As linhas proibidas, especialmente as de [OIII], [NII] e [SII], também são sensíveis à temperatura e pressão, portanto, definir uma razão entre as intensidades das linhas [OIII] de 436,3 nm e 500,7 nm permite aos astrônomos determinar a densidade dos elétrons e temperatura de um corpo celeste.

\subsection{Galáxias}

As galáxias são enormes estruturas que abrigam os mais variados tipos de estrelas, nebulosas, supernovas e até quasares, dessa forma, é lógico e correto supor que o espectro de uma galáxia seja a média de todas as emissões de todos os objetos que a compõem \cite{almeida}. Portanto, ao contrário de uma estrela, com seu espectro como uma curva de corpo negro e linhas de absorção, o espectro de uma galáxia tem como característica uma abundância de minúsculos picos e vales, correspondendo ao somatório dos espectros de emissão e absorção dos mais diversos elementos químicos em seus mais diversos graus de ionização.

\begin{figure}[h!]
  \caption{\textbf{Dados espectroscópicos de uma galáxia lenticular e uma galáxia \textit{starburst}}}\label{galaxy_spectra}
	\begin{subfigure}{0.5\textwidth}
		\centering
		\includegraphics[width=.95\linewidth]{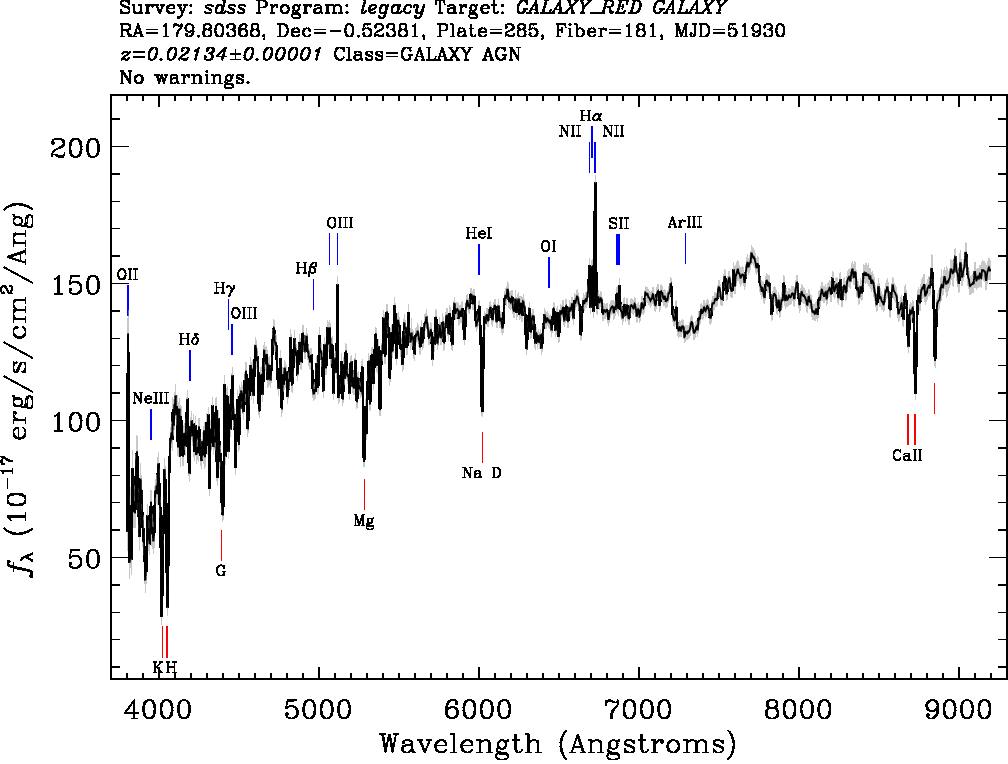}
		\caption{Galáxia lenticular IC 753.}\label{galaxy_lenticular}
	\end{subfigure}
	\begin{subfigure}{0.5\textwidth}
		\centering
		\includegraphics[width=.95\linewidth]{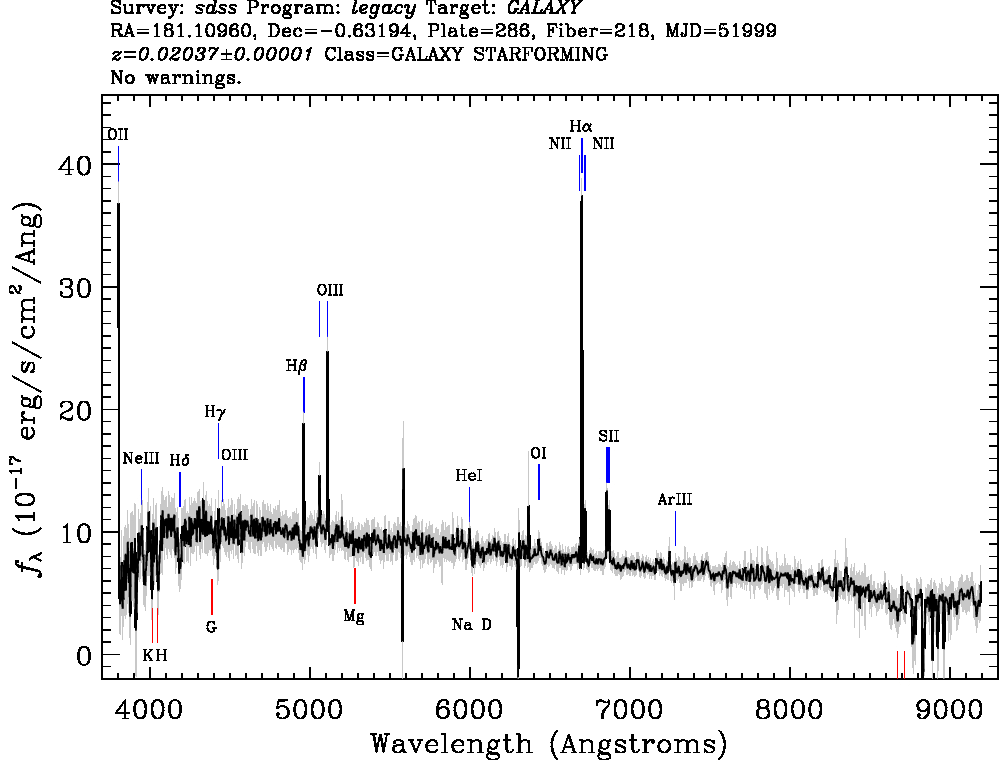}
		\caption{Galáxia LEDA 1138516.}\label{galaxy_starburst}
	\end{subfigure}
	\caption*{Fonte: \textcite{sdss}}
\end{figure}

O espectro de uma galáxia mostra diversos picos referentes à emissão de radiação proveniente de quasares, regiões H II, nebulosas planetárias e remanescentes de supernovas. Diferenças nos espectros podem indicar se a galáxia é composta por estrelas antigas, ou seja, seu espectro é avermelhado e de alta metalicidade, ou se é uma galáxia \textit{starburst}, com alta taxa de formação de novas estrelas. Galáxias antigas (Figura \ref{galaxy_lenticular}) têm seu espectro avermelhado, rico em metais e com menor índice de emissões de hidrogênio ou oxigênio ionizado devido à baixa densidade de regiões H II e ``a maior parte de sua luz vem de gigantes vermelhas enquanto a maior parte de sua massa reside em estrelas da sequência principal de baixa massa'' \cite[p.~404]{karttunen}. Galáxias jovens e com alta taxa de formação de estrelas (Figura \ref{galaxy_starburst}), chamadas galáxias \textit{starburst}, contém muitas regiões H II e estrelas jovens, logo, o espectro geral da galáxia é mais azulado e contém muitos picos referentes às emissões de hidrogênio, oxigênio e outros elementos comuns em nuvens moleculares e protoestrelas.

A medição do desvio Doppler dos espectros também pode ser usada para determinar a velocidade da rotação de uma galáxia. Por estar em rotação, um lado da galáxia espiral está se aproximando, enquanto o outro está se afastando. Determinar a diferença nos desvios das linhas espectrais de cada um dos lados de uma galáxia permite calcular sua velocidade de rotação.

\subsection{Quasares}

O nascimento e evolução da radioastronomia na década de 1950 fez com que muitos astrônomos detectassem fontes pontuais e intensas em ondas de rádio. Ao direcionar telescópios ópticos para observar como uma dessas fontes (3C 273) se parecia no espectro visível, foi notado que a aparência era muito similar à de uma estrela (Figura \ref{example_visible_quasar}). No entanto, ao realizar a espectrometria notou-se que sua emissão em nada se parecia com uma estrela. Esse objeto de aparência quase estelar foi chamados de \textit{quasar} (do inglês, \textit{quasi-stellar radio source})\footnote{Os astrofísicos preferem a nomenclatura \textit{QSO} (de \textit{quasi-stellar object}), afinal, nem todos os quasares emitem sinais de rádio.}.

\begin{figure}[h!]
  \centering
  \caption{\textbf{Espectro do quasar SDSS J115945.79-003647.0}}\label{qso}
  \includegraphics[width=.7\textwidth]{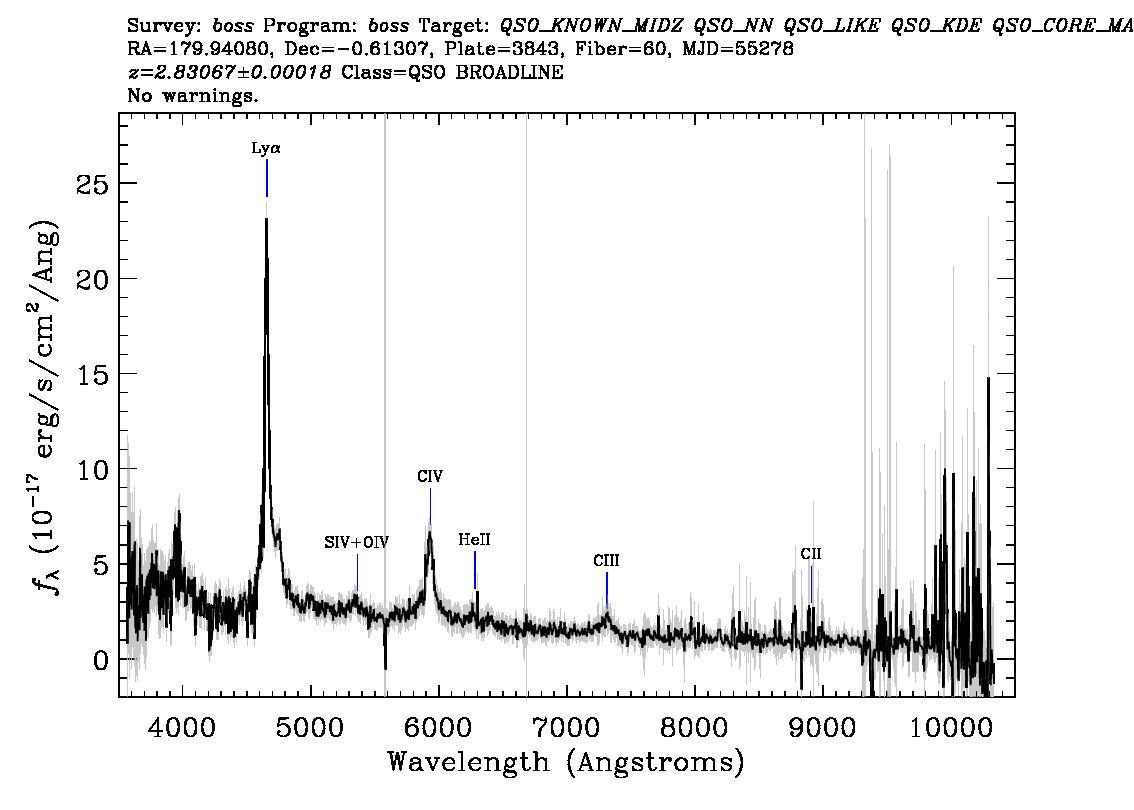}
  \caption*{Fonte: \textcite{sdss}}
\end{figure}

Posteriormente, em 1963, Maarten Schmidt percebeu que o espectro incomum dos quasares devia-se ao fato de estarem se movendo a velocidades relativísticas, em frações consideráveis da velocidade da luz, causando um enorme desvio para o vermelho. O espectro da galáxia 3C 48, segundo \textcite{carroll}, mostra uma velocidade radial de $z=0,367$, o que representa 30\% da velocidade da luz. Tais desvios, típicos dos quasares, fazem com que as emissões de alta energia no espectro do ultravioleta tornem-se visíveis no espectro óptico, motivo pela qual os quasares são extremamente brilhantes, mesmo em longas distâncias. Então uma característica dos espectros dos quasares são que as fortes emissões de Ly-α (série de Lyman), MgII e NV, antes no ultravioleta, estejam deslocadas para o espectro visível, ou comprimentos ainda mais longos, como o infravermelho-próximo (Figura \ref{qso}).

\begin{figure}[h!]
  \centering
  \caption{\textbf{Espectro do quasar PSO J006.1240+39.2219}}\label{pso}
  \includegraphics[width=.7\textwidth]{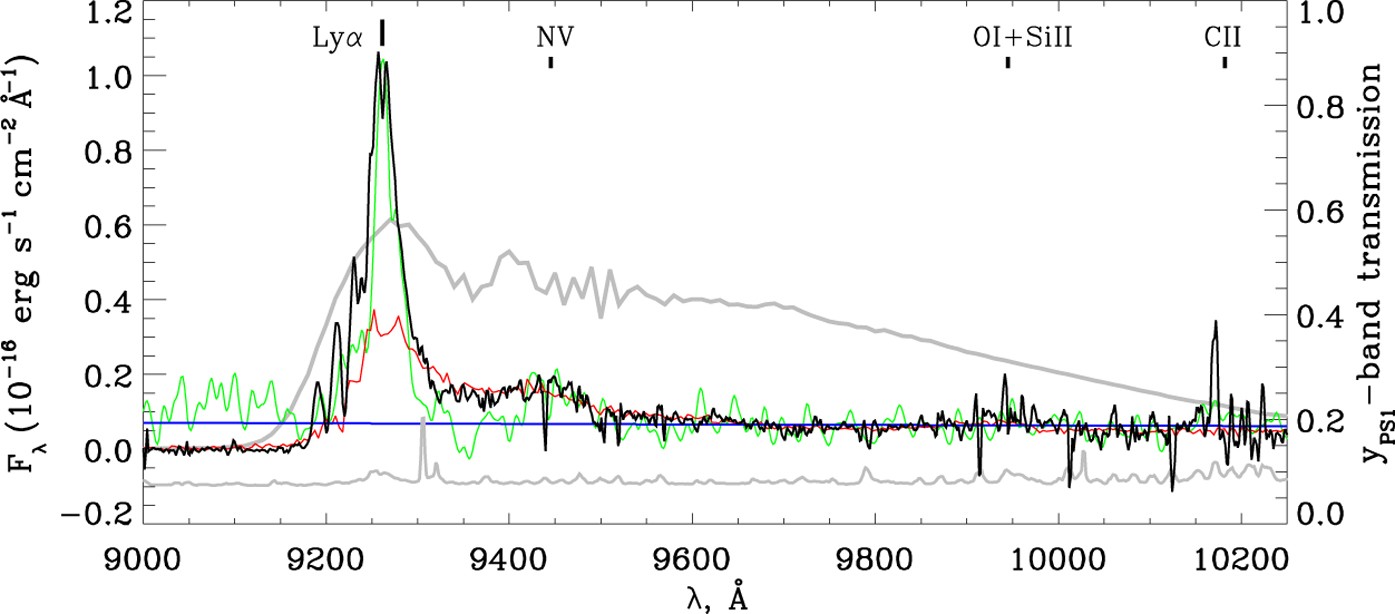}
  \caption*{Fonte: \textcite{koptelova}}
\end{figure}

A Figura \ref{pso} mostra um caso extremo do espectro de um quasar altamente luminoso em Ly-α, cuja emissão normalmente se dá em 1216 Å, no ultravioleta. O gráfico mostra que o efeito Doppler relativístico deslocou a linha de Ly-α para entre 9200 e 9300 Å, no infravermelho-próximo, cujas medições de \textcite{koptelova} mostram um desvio de $z=6,62$. Essas medições de enormes deslocamentos Doppler das linhas de emissão contribuem como uma das comprovações da expansão do Universo \cite{kepler,horvath:altasenergias} e, consequentemente, da teoria do Big Bang.

Essas galáxias extremamente luminosas têm suas emissões como um corpo negro, proveniente das estrelas que a compõem, mas emitem picos de um espectro não térmico, ou seja, cuja fonte não são estrelas, mas um núcleo galáctico ativo (ou AGN). Essas observações foram associadas com outras galáxias com fortes emissões de rádio e agrupadas como \textit{galáxias ativas} \cite{kepler,horvath:altasenergias}. Hoje, é aceito que grande parte da energia de um AGN provém de um buraco negro supermassivo localizado em seu centro que, ao acretar matéria, emite enormes quantidades de radiação à medida que a matéria se acelera e espirala no disco de acreção.

\subsection{Planetas, satélites e cometas do Sistema Solar}

A espectroscopia astronômica não é útil somente na pesquisa extrassolar. Ainda há muito a ser compreendido sobre o nosso próprio Sistema Solar. Como as viagens espaciais tripuladas ainda são limitadas e de risco elevado, os astrônomos recorrem à espectroscopia de sondas e rovers para determinar a composição do solo e da atmosfera de outros corpos celestes, como Vênus, Marte ou Titã, por exemplo.

Como planetas, satélites e cometas são bem mais frios do que estrelas, não há espectro de emissão. Seus espectros são compostos por absorções moleculares da luz refletida do Sol e realizados predominantemente no infravermelho — por isso a importância de telescópios infravermelho, como o WISE e o James Webb. As linhas de absorção moleculares dão evidências da composição de sua superfície ou atmosfera \cite{shaw}. No caso de rovers, existem técnicas de análise espectroscópica envolvendo a vaporização de materiais com lasers que serão detalhadas nas próximas seções.

Em termos de exploração espacial, além dos planetas, existem dois tipos de corpos celestes que chamam uma atenção especial: os satélites naturais e os cometas.

\subsubsection{Cometas}

\begin{figure}[h!]
  \centering
  \caption{\textbf{Espectro do cometa Hyakutake}}\label{comet}
  \includegraphics[width=.75\textwidth]{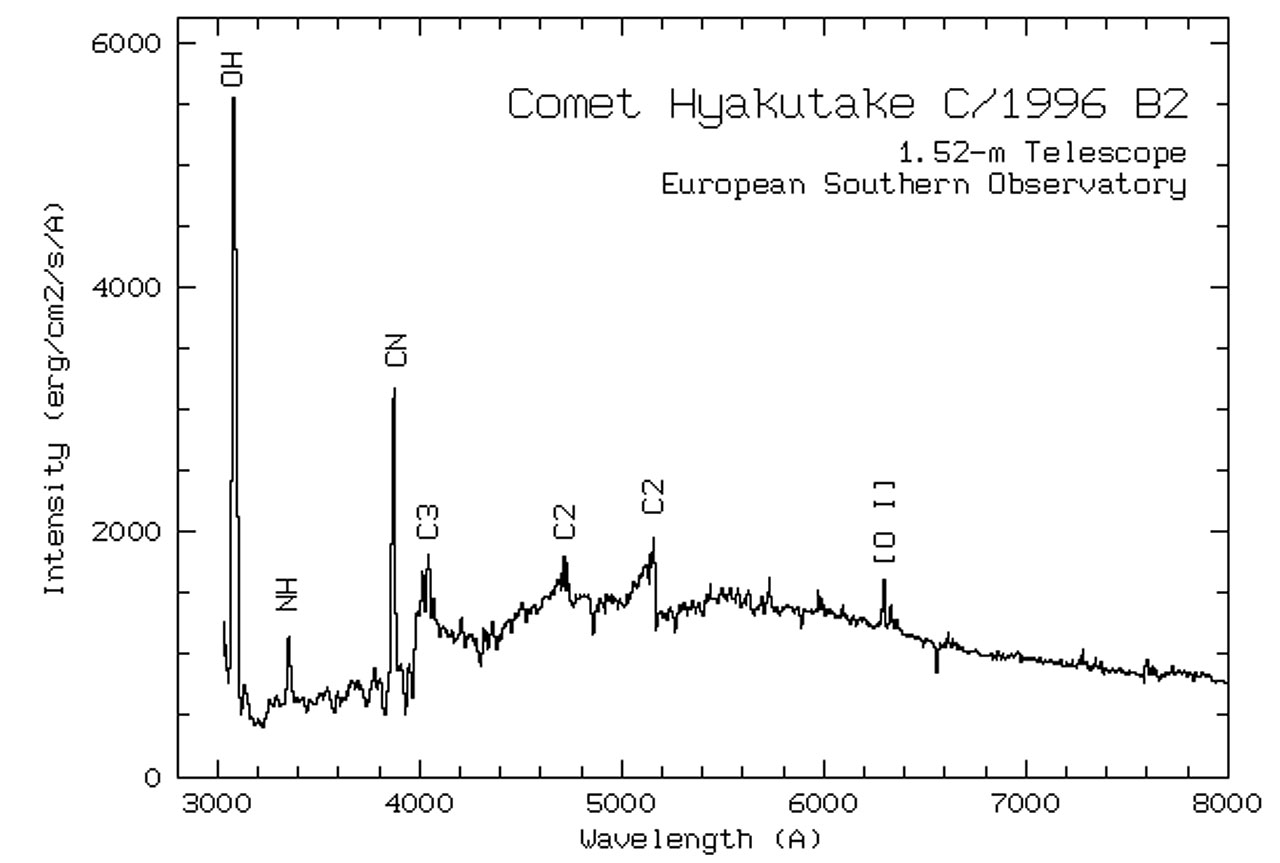}
  \caption*{Fonte: \textcite{eso1996}}
\end{figure}

Os cometas têm uma órbita altamente elíptica e seus núcleos são feitos primariamente de gelo e compostos de carbono, nitrogênio e hidrogênio \cite{carroll}. A existência desses compostos é comprovada pelos seus espectros. De acordo com \textcite{hyland}, o espectro dos cometas é composto por linhas de emissão brilhantes de seus compostos — principalmente carbono molecular, $\mathrm{NH}$, $\mathrm{CN}$ e $\mathrm{NH_2}$ — sobreposto em um fraco espectro contínuo devido ao reflexo da luz do Sol (Figura \ref{comet}).

As tonalidades de cores dos cometas são, portanto, indicativos de sua composição devido ao espectro de emissão desses compostos. A cauda de íons de um cometa é azul, pois os íons $\mathrm{CO^+}$ absorvem e irradiam fótons em comprimento de onda aproximado de 420 nm \cite{carroll}. A fotodissociação da molécula diatômica do carbono ($\mathrm{C_2}$), por exemplo, é responsável pela emissão de cor verde na coma dos cometas \cite{borsovszky}.

\subsubsection{Titã}

\begin{figure}[h!]
  \centering
  \caption{\textbf{Espectro da atmosfera de Titã obtida pela sonda Voyager 1}}\label{titan}
  \includegraphics[width=.75\textwidth]{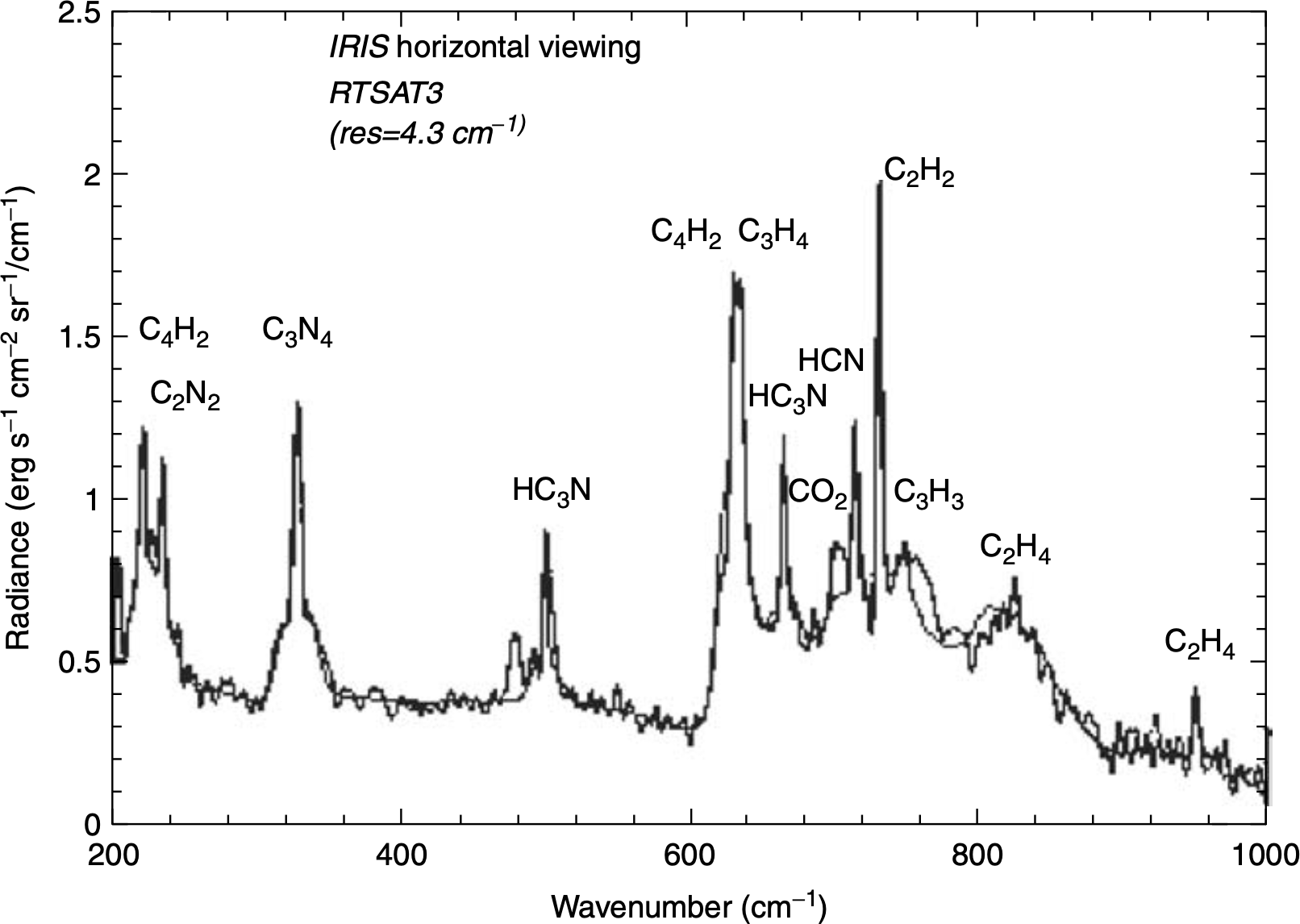}
  \caption*{Fonte: \textcite{shaw}}
\end{figure}

Com 40\% do diâmetro da Terra, Titã é o maior satélite de Saturno, e o segundo maior satélite do Sistema Solar, atrás apenas de Ganimedes. As análises dos espectros de Titã obtidos pelo instrumento IRIS da Voyager 1 (Figura \ref{titan}) e da missão Cassini-Huygens mostram que sua atmosfera é composta por aproximadamente 90\% de nitrogênio, e 1,7–4,5\% de hidrogênio e hidrocarbonetos, principalmente metano ($\mathrm{CH_4}$), etileno ($\mathrm{C_2H_4}$) e acetileno ($\mathrm{C_2H_2}$), como mostra a Figura \ref{titan} a seguir \cite{figueiredo,shaw}.

Segundo \textcite{figueiredo}, além de uma atmosfera densa, Titã possui um ciclo de metano muito similar ao ciclo da água da Terra, com nuvens, precipitação, rios e mares de metano e outros hidrocarbonetos. A existência de moléculas complexas de hidrocarbonetos, somado com o interior aquecido de Titã devido ao decaimento radioativo, também desperta o interesse de astrobiólogos especialistas em química prebiótica, que levantam a hipótese da existência de formas de vida primordiais \cite{shaw}.

\subsubsection{Lua terrestre}

\begin{figure}[h!]
  \centering
  \caption{\textbf{Espectro do solo lunar obtido pelo Chandrayaan-3}}\label{moon}
  \includegraphics[width=.75\textwidth]{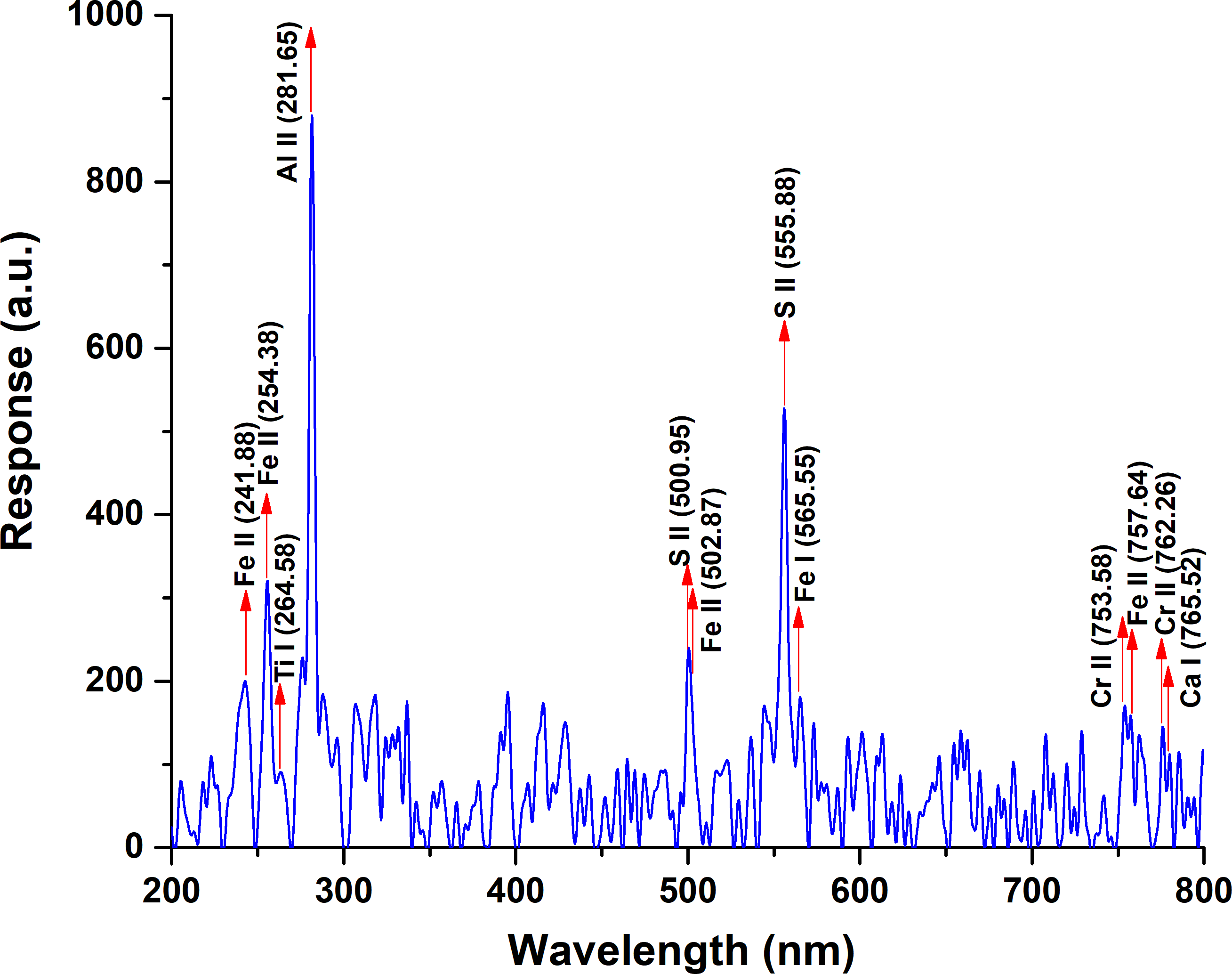}
  \caption*{Fonte: \textcite{isro}}
\end{figure}

Diversas expedições retornaram com amostras de rocha e poeira da superfície da Lua, permitindo aos cientistas determinar a composição do solo do nosso satélite natural em laboratório. No entanto, para pesquisas de campo operadas remotamente (como por meio de rovers), recorre-se à técnica da espectroscopia. A pesquisa mais recente, realizada pelo rover Chandrayaan-3, da ISRO, a agência espacial indiana, mediu a composição do solo lunar \textit{in situ} próximo ao polo sul. As análises preliminares divulgadas em agosto de 2023 (Figura \ref{moon}) mostram a presença de alumínio, enxofre, cálcio, ferro, cromo e titânio.

O Chandrayaan-3 realiza a espectroscopia do solo lunar por meio de uma técnica chamada LIBS, do inglês, \textit{Laser-Induced Breakdown Spectroscopy} \cite{isro}. Pulsos de laser altamente energéticos na superfície de uma amostra, como o solo lunar, geram vaporização e plasma em altas temperaturas, na ordem de 10.000 a 20.000 K, em uma câmara de ablação. A emissão de luz do material vaporizado ao retornar para seu estado fundamental é coletada, dispersada e medida por um espectrômetro \cite{costa}.

\section{Conclusão}

Considerando a relevância crucial da espectroscopia para o progresso da nossa compreensão dos objetos e fenômenos astronômicos, associada com a necessidade da abordagem desses temas de acordo com as competências e habilidades da BNCC (Base Nacional Comum Curricular), sugere-se uma maior ênfase e inclusão deste tema nos currículos educacionais. Uma incorporação mais sistemática da teoria espectroscópica e sua aplicação em contextos acadêmicos reais pode servir para reforçar o conhecimento fundamental para futuros pesquisadores e profissionais da área. Essa abordagem não somente fortaleceria a compreensão da análise de dados entre os cientistas em desenvolvimento durante o Ensino Superior, mas também reforçaria a aplicabilidade dos conceitos teóricos da Física Moderna em projetos práticos voltados para a pesquisa científica para estudantes do Ensino Médio. Portanto, este estudo defende uma abordagem educacional mais aprofundada sobre esses conceitos fundamentais da natureza da matéria, assegurando que a sua aplicabilidade seja reconhecida e destacada.

A complexidade da investigação dos espectros de emissão e absorção discutida ao longo deste artigo evidencia a importância da espectroscopia no campo da astrofísica. Os variados espectros dos diversos tipos de corpos celestes não apenas desvendam as suas características inerentes, mas também servem como uma ferramenta analítica robusta que proporciona uma exploração minuciosa das suas propriedades físicas e químicas intrínsecas. Esse exame detalhado permite a determinação da composição, estado térmico, densidade, massa e movimento relativo de objetos celestes, afirmando o papel imperativo da espectroscopia na pesquisa em Astrofísica.

Após uma análise aprofundada de uma grande variedade de espectros celestes, foram identificadas diferenças significativas devido às razões distintas, demonstrando a capacidade dos métodos espectroscópicos de decifrar fenômenos astrofísicos complexos e muito distantes. Apesar de sua relevância evidente na pesquisa científica, a espectroscopia e, consequentemente, o espectro eletromagnético e os conceitos fundamentais do átomo de Bohr e das transições de nível de energia, são, às vezes, esquecidos em um lugar secundário nos ambientes educacionais, especialmente nos níveis do Ensino Médio.

A análise dos espectros de diversos corpos celestes e seu sucesso em explicar os diversos fenômenos astronômicos sustenta a tese de que o domínio da espectroscopia é uma metodologia indispensável para a pesquisa, merecendo um lugar de maior destaque no discurso acadêmico e na pedagogia.

\printbibliography

@book{kepler,
    title = {Astronomia \& Astrofísica},
    author = {K. S. Souza Oliveira~Filho and M. F. O. Saraiva},
    year = {2017},
    edition = {4},
    publisher = {Livraria da Física},
    location = {São Paulo}
}

@book{gallaway,
	title = {An Introduction to Observational Astrophysics},
	author = {M. Gallaway},
	year = {2020},
	edition = {2},
	location = {Berlim},
	publisher = {Springer}
}

@book{karttunen,
	title = {Fundamental Astronomy},
	author = {Hannu Karttunen and Pekka Kröger and Heikki Oja and Markku Poutanen and Karl Johan Donner},
	year = {2017},
	edition = {6},
	location = {Nova York},
	publisher = {Springer-Verlag Berlin Heidelberg}
}

@book{kitchin,
	title = {Astrophysical Techniques},
	author = {C. R. Kitchin},
	year = {2020},
	edition = {7},
	publisher = {CRC Press},
	location = {Boca Raton}
}

@book{horvath:estelar,
	title = {Fundamentos da evolução estelar, supernovas e objetos compactos},
	author = {Jorge E. Horvath},
	location = {São Paulo},
	publisher = {Livraria da Física},
	year = {2011}
}

@book{horvath:altasenergias,
	title = {Astrofísica de Altas Energias},
	subtitle = {Uma Premiére},
	author = {Jorge E. Horvath},
	location = {São Paulo},
	publisher = {Editora da Universidade de São Paulo},
	year = {2020}
}

@book{carroll,
	title = {An Introduction to Modern Astrophysics},
	author = {B. W. Carroll and D. A. Ostlie},
	edition = {2},
	location = {Cambridge},
	publisher = {Cambridge University Press},
	year = {2017}
}

@article{branchesi,
	title = {Multi-messenger astronomy: gravitational waves, neutrinos, photons and cosmic rays},
	author = {M. Branchesi},
	journaltitle = {Journal of Physics: Conference Series},
	volume = {718},
	number = {2},
	year = {2016},
	doi = {10.1088/1742-6596/718/2/022004}
}

@book{lena,
	title = {Observational Astrophysics},
	author = {P. Léna and D. Rouan and F. Lebrun and F. Mignard and D. Pelat},
	edition = {3},
	publisher = {Springer},
	location = {Berlim},
	year = {2012}
}

@incollection{silva:mclean,
	title = {Introduction to Telescopes},
	author = {D. Silva and I. S. McLean},
	pages = {2-42},
	maintitle = {Planets, Stars and Stellar Systems},
	editor = {T. D. Oswalt and I. S. McLean},
	location = {Berlim},
	publisher = {Springer Dordrecht},
	volume = {1},
	year = {2013}
}

@article{pagliarini:almeida,
	title = {Leituras por alunos do Ensino Médio de textos de cientistas sobre o início da física quântica},
	author = {C. R. Pagliarini and M. J. P. M. Almeida},
	journaltitle = {Ciência \& Educação},
	volume = {22},
	number = {2},
	pages = {299-317},
	year = {2016},
	doi = {10.1590/1516-731320160020003}
}

@article{silva:almeida,
	title = {Física quântica no Ensino Médio: o que dizem as pesquisas},
	author = {A. C. Silva and M. J. P. M. Almeida},
	journaltitle = {Caderno Brasileiro de Ensino de Física},
	volume = {28},
	number = {3},
	pages = {264-652},
	year = {2011}
}

@inproceedings{sanches,
	title = {A inserção da Física Moderna e Contemporânea no Ensino Médio},
	author = {M. B. Sanches and J. H. L. Oliveira and M. C. D. Neves and S. O. Resquetti},
	booktitle = {Atlas do X Encontro de Pesquisa em Ensino de Física},
	location = {Londrina},
	publisher = {Sociedade Brasileira de Física},
	year = {2006}
}

@inproceedings{carvalho,
	title = {Ciência e arte, razão e imaginação: complementos necessários à compreensão da Física Moderna},
	author = {S. H. M. Carvalho and J. Zanetic},
	booktitle = {Anais do IX Encontro de Pesquisa em Ensino de Física},
	location = {Jaboticatubas},
	publisher = {Sociedade Brasileira de Física},
	year = {2004}
}

@article{ostermann,
	title = {Uma revisão bibliográfica sobre a área de pesquisa ``Física moderna e contemporânea no Ensino Médio''},
	author = {F. Ostermann and M. A. Moreira},
	journaltitle = {Investigações em Ensino de Ciências},
	volume = {5},
	pages = {23-48},
	year = {2000}
}

@book{oliveira,
	title = {Introdução ao Eletromagnetismo},
	author = {I. Oliveira},
	location = {São Paulo},
	publisher = {Blucher},
	year = {2021}
}

@book{halliday,
	title = {Fundamentos de Física},
	subtitle = {Óptica e Física Moderna},
	author = {D. Halliday and R. Resnick and J. Walker},
	edition = {10},
	volume = {4},
	location = {Rio de Janeiro},
	publisher = {LTC},
	year = {2019}
}

@article{kyrola,
	title = {Retrieval of atmospheric parameters from GOMOS data},
	author = {E. Kyrölä and others},
	journaltitle = {Atmospheric Chemistry and Physics},
	volume = {10},
	number = {23},
	pages = {118811-11903},
	year = {2010},
	doi = {10.5194/acp-10-11881-2010}
}

@inproceedings{fraunhofer,
	title = {Bestimmung des Brechungs- und des Farben-Zerstreuungs – Vermögens verschiedener Glasarten, in Bezug auf die Vervollkommnung achromatischer Fernröhre},
	author = {Joseph von Fraunhofer},
	booktitle = {Denkschriften der Königlichen Akademie der Wissenschaften zu München},
	publisher = {Bayerische Akademie der Wissenschaften},
	volume = {5},
	pages = {193-226},
	location = {Munique},
	year = {1817},
	url = {https://books.google.com.br/books?id=2-AAAAAAYAAJ}
}

@book{eisberg,
	title = {Física Quântica},
	subtitle = {Átomos, Moléculas, Sólidos, Núcleos e Partículas},
	author = {R. Eisberg and R. Resnick},
	location = {Rio de Janeiro},
	publisher = {Elsevier},
	year = {1979}
}

@inproceedings{sdss,
	author = {{Kollmeier}, Juna and {Anderson}, S.~F. and {Blanc}, G.~A. and {Blanton}, M.~R. and {Covey}, K.~R. and {Crane}, J. and {Drory}, N. and {Frinchaboy}, P.~M. and {Froning}, C.~S. and {Johnson}, J.~A. and {Kneib}, J. -P. and {Kreckel}, K. and {Merloni}, A. and {Pellegrini}, E.~W. and {Pogge}, R.~W. and {Ramirez}, S.~V. and {Rix}, H.~W. and {Sayres}, C. and {S{\'a}nchez-Gallego}, Jos{\'e} and {Shen}, Yue and {Tkachenko}, A. and {Trump}, J.~R. and {Tuttle}, S.~E. and {Weijmans}, A. and {Zasowski}, G. and {Barbuy}, B. and {Beaton}, R.~L. and {Bergemann}, M. and {Bochanski}, J.~J. and {Brandt}, W.~N. and {Casey}, A.~R. and {Cherinka}, B.~A. and {Eracleous}, M. and {Fan}, X. and {Garc{\'\i}a}, R.~A. and {Green}, P.~J. and {Hekker}, S. and {Lane}, R.~R. and {Longa-Pe{\~n}a}, P. and {Mathur}, S. and {Meza}, A. and {Minchev}, I. and {Myers}, A.~D. and {Nidever}, D.~L. and {Nitschelm}, C. and {O'Connell}, J.~E. and {Price-Whelan}, A.~M. and {Raddick}, M.~J. and {Rossi}, G. and {Sankrit}, R. and {Simon}, J.~D. and {Stutz}, A.~M. and {Ting}, Y. -S. and {Trakhtenbrot}, B. and {Weaver}, B.~A. and {Willmer}, C.~N.~A. and {Weinberg}, D.~H.},
	title = "{SDSS-V Pioneering Panoptic Spectroscopy}",
	booktitle = {Bulletin of the American Astronomical Society},
	year = 2019,
	volume = {51},
	month = sep,
	eid = {274},
	pages = {274},
	adsurl = {https://ui.adsabs.harvard.edu/abs/2019BAAS...51g.274K}
}

@book{leblanc,
	title = {An Introduction to Stellar Astrophysics},
	author = {Francis LeBlanc},
	location = {Chichester},
	publisher = {John Wiley and Sons},
	year = {2010}
}

@article {almeida,
	title = {Qualitative Interpretation of Galaxy Spectra},
	author = {J. Sánchez Almeida and R. Terlevich and E. Terlevich and R. Cid Fernandes and A. B. Morales-Luis},
	journaltitle = {The Astrophysical Journal},
	publisher = {The American Astronomical Society},
	volume = {756},
	number = {2},
	year = {2012},
	doi = {10.1088/0004-637X/756/2/163}
}

@article{peimbert,
	title = {Nebular Spectroscopy: A Guide on HII Regions and Planetary Nebulae},
	author = {Manuel Peimbert and Antonio Peimbert and Gloria Delgado-Inglada},
	journaltitle = {Publications of the Astronomical Society of the Pacific},
	volume = {129},
	number = {978},
	year = {2017},
	doi = {10.1088/1538-3873/aa72c3}
}

@article{tosi,
	title = {Energia nuclear vê hora de superar tabus e virar opção estratégica para o Brasil},
	author = {Marcos Tosi},
	journaltitle = {Gazeta do Povo},
	date = {2023-08-12},
	url = {https://www.gazetadopovo.com.br/economia/energia-nuclear-ve-hora-de-superar-tabus-e-salvar-o-planeta},
	urldate = {2023-08-31}
}

@article{koptelova,
	title = {Discovery of a very Lyman-$\alpha$-luminous quasar at z = 6.62},
	author = {Koptelova, Ekaterina and Hwang, Chorng-Yuan and Yu, Po-Chieh and Chen, Wen-Ping and Guo, Jhen-Kuei},
	date = {2017-02-02},
	doi = {10.1038/srep41617},
	journal = {Scientific Reports},
	number = {1},
	pages = {41617},
	volume = {7},
	year = {2017}
}

@article{rocha,
	title = {A Era Dourada dos Árabes},
	author = {Hélio Jacques Rocha-Pinto},
	journal = {Revista Brasileira de Astronomia},
	volume = {5},
	number = {18},
	year = {2023}
}

@book{bless,
	title = {Discovering the Cosmos},
	author = {Robert C. Bless},
	publisher = {University Science Books},
	location = {Sausalito},
	year = {1996}
}

@inproceedings{magrini,
	address = {Berlin, Heidelberg},
	author = {Magrini, Laura and Leisy, Pierre and Corradi, Romano L.M. and Perinotto, Mario and Mampaso, Antonio and V{\'\i}lchez, Jos{\'e}},
	booktitle = {Planetary Nebulae Beyond the Milky Way},
	editor = {Stanghellini, L. and Walsh, J. R. and Douglas, N. G.},
	pages = {247-251},
	publisher = {Springer Berlin Heidelberg},
	title = {Spectroscopy of Planetary Nebulae in Sextans A and Sextans B},
	year = {2006}
}

@book{shaw,
	title = {Astrochemistry},
	subtitle = {from Astronomy to Astrobiology},
	author = {Andrew M. Shaw},
	publisher = {Wiley},
	location = {Chichester},
	year = {2006}
}

@article{borsovszky,
	title = {Photodissociation of dicarbon: How nature breaks an unusual multiple bond},
	author = {Jasmin Borsovszky and Klaas Nauta and Jun Jiang and Christopher S. Hansen and Laura K. McKemmish and Robert W. Field and John F. Stanton and Scott H. Kable and Timothy W. Schmidt},
	journal = {PNAS},
	volume = {118},
	number = {52},
	year = {2021}
}

@online{eso1996,
	author = {ESO},
	title = {Spectrum of Comet Hyakutake (3000-8000 A)},
	url = {https://www.eso.org/public/spain/news/eso9620},
	year = {1996},
	urldate = {2023-10-10}
}

@article{hyland,
	title = {Near-UV and optical spectroscopy of comets using the ISIS spectrograph on the WHT},
	author = {M. G. Hyland and A. Fitzsimmons and C. Snodgrass},
	journal = {Monthly Notices of the Royal Astronomical Society},
	volume = {484},
	number = {1},
	year = {2019}
}

@inbook{figueiredo,
	title = {Luas geladas do Sistema Solar},
	author = {Douglas Borges de Figueiredo},
	pages = {235},
	maintitle = {Astrobiologia, uma ciência emergente},
	bookauthor = {Douglas Galante and Evandro Pereira da Silva and Fábio Rodrigues and Jorge E. Horvath and Márcio G. B. de Avellar},
	editortype = {organizer},
	location = {São Paulo},
	publisher = {IAG/USP},
	year = {2016}
}

@online{isro,
	author = {ISRO},
	title = {LIBS confirms the presence of Sulphur (S) on the lunar surface through unambiguous in-situ measurements},
	url = {https://www.isro.gov.in/LIBSResults.html},
	date = {2023-08-28},
	urldate = {2023-11-05}
}

@article{costa,
	title = {Laser-Induced Breakdown Spectroscopy (LIBS): Histórico, Fundamentos, Aplicações e Potencialidades},
	author = {Vinicius C. Costa and Amanda S. Augusto and Jeyne P. Castro and Raquel C. Machado and Daniel F. Andrade and Diego V. Babos and Marco A. Sperança and Raimundo R. Gamela and Edenir R. Pereira-Filho},
	journal = {Química Nova},
	volume = {42},
	number = {5},
	pages = {527-545},
	year = {2019}
}

\end{document}